\documentclass{article}

\usepackage{xcolor}
\usepackage{subfig}
\usepackage{graphicx}
\usepackage[nonatbib,final]{tccml_neurips_2020}
\usepackage[utf8]{inputenc} 
\usepackage[T1]{fontenc}    
\usepackage{hyperref}       
\usepackage{url}            
\usepackage{booktabs}       
\usepackage{amsfonts}       
\usepackage{nicefrac}       
\usepackage{microtype}      

\title{Loosely Conditioned Emulation of Global Climate Models With Generative Adversarial Networks}

\author{
  Alexis Ayala$^1$, Christopher Drazic$^1$, Brian Hutchinson$^{1,2}$, Ben Kravitz$^3$, and Claudia Tebaldi$^4$\\
  $^1$ Computer Science Department, Western Washington University, Bellingham, WA\\
  $^2$ Computing \& Analytics Division, Pacific Northwest National Laboratory, Richland, WA\\
  $^3$ Earth and Atmospheric Sciences Department, Indiana University, Bloomington, IN\\
  $^4$ Joint Global Change Research Institute, Pacific Northwest National Laboratory, College Park, MD}

\begin{document}

\maketitle

\section{Introduction}

Climate models encapsulate our best understanding of the Earth system, allowing research to be conducted on its future under alternative assumptions of how human-driven climate forces are going to evolve.  An important application of climate models is to provide metrics of mean and extreme climate changes, particularly under these alternative future scenarios, as these quantities drive the impacts of climate on society and natural systems \cite{Moraetal2018, Forzierietal2018, Raymond2020}.  Furthermore, efforts in integrated modeling seek to ``close the loop,'' by having impacts on society  feedback on societal conditions that drive emissions \cite{Calvin_2018}. Because of the need to explore a wide range of alternative scenarios and other sources of uncertainties in a computationally efficient manner, climate models can only take us so far, as they require large computational resources, with a single simulation of the 21$^{\mathrm{st}}$ century taking on the order of weeks on a supercomputer. 
The computational requirements expand considerably when attempting to characterize extreme events, which are rare and thus demand long and numerous simulations exploring a noisy system in order to accurately represent their changing statistics.

Climate model emulators address some of these problems.  Trained on climate model output, emulators are simpler, data-driven tools (often obtained through parametric fits like regressions) that are less accurate or less complex than climate models but can produce values in fractions of a second on a laptop. Their computational cost, when significant, is made upfront in the training phase.  Traditionally, emulators like Pattern Scaling  have been used to approximate average quantities, like annual or seasonal or monthly average temperature and precipitation \cite{Santeretal1990,Castruccio2014, Holden2014,Tebaldi2014, Hergeretal2015,kravitzetal2017,Linketal2019, Beusch2020}. Recently the accuracy of some of these techniques for representing extremes has been documented as well \cite{Tebaldi_2020}. ``Top-down'' approaches to emulation involve directly approximating metrics themselves, like the hottest or the wettest day of the year. Alternatively, a ``bottom-up'' approach tackles the emulation of the building blocks (in these cases, daily temperatures and precipitation during the year) and then compute the metric of interest. Examples include stochastic weather generators \cite{Semenov1997, Kilsby2007,Fatichi2011}, which rely on parametrizing the distribution of the weather variable and randomly sample its realizations. Weather generators have been usually developed for limited domains and specific applications, rarely facing the issue of representing non-stationarities and non-linearities, which are critical for integrated modeling of impacts,  arbitrary domains, and  scenario generation.

Increasingly, joint efforts between climate science and machine learning are being formed to tackle some of the most complex data-driven problems~\cite{Reichstein2019,Rolnick2019}. So far most of the applications have focused on bringing deep learning in aid of better model forecasts, model parameterizations, or in substitution of climate models~\cite{Cohen2019,Grover2015,Ham2019,Shi2017,Jones2017,He2020,Schneider2017,Weber2020,schmidt2020modeling}; of better detection of signals, from extreme events to large scale patterns of anthropogenic changes amidst the internal noise of the climate system~\cite{Liu2016,Wang2019,Klemmer2019,Barnes2020,Toms2020,Wills2020}; and of spatial in-filling in the case of fine-scale features that models would be too expensive, or plainly unable, to generate, or observations cannot cover~\cite{Kuhnlein2014,Amato2020,Vandal17,Stengel2020adversarial}.

Here we use deep learning in a proof of concept that lays the foundation for  emulating global climate model output for different scenarios.  We train Generative Adversarial Networks (GANs) that emulate daily precipitation output from a fully coupled Earth system model. Our GANs are trained to produce samples in the form of $T \times H \times W$ tensors, where $T$ denotes the number of timesteps (days) and $H$ and $W$ are the spatial height and width, respectively, of a regular grid discretizing the globe. The goal is for these samples to be statistically indistinguishable from samples of the same dimension drawn from a state-of-the-art earth system model.  
Our trained GAN can rapidly generate numerous realizations at a vastly reduced computational expense,  
compared to large ensembles of climate models \cite{Kay2015, Lehner2020}, which greatly aids in estimating the statistics of extreme events. Compared to our prior ``DeepClimGAN'' \cite{puchko2020}, we find that the approach proposed here produces significantly higher quality spatio-temporal samples. 

\section{Model}
\begin{figure}
    \centering
    \subfloat[Generator]{\includegraphics[width=0.35\linewidth]{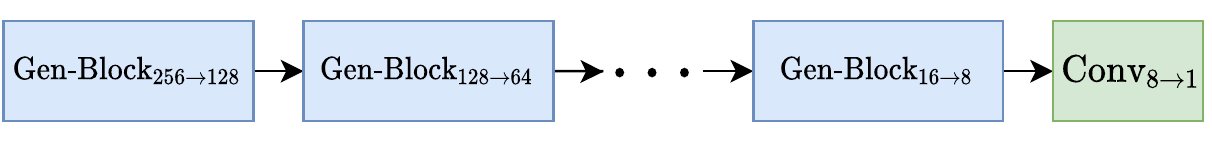}}
    \hspace{0.5cm}
    \subfloat[Critic]{\includegraphics[width=0.55\linewidth]{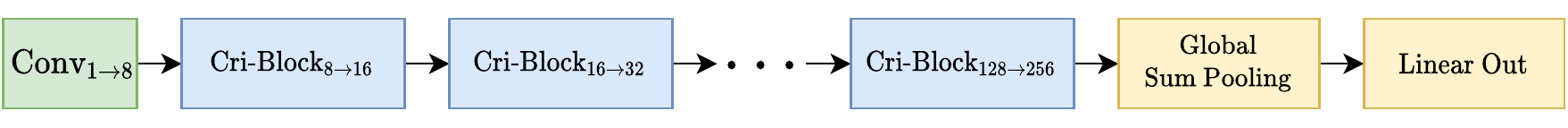}}\\
    \vspace{-5pt}
    \subfloat[Generator block.]{\includegraphics[width=0.45\linewidth]{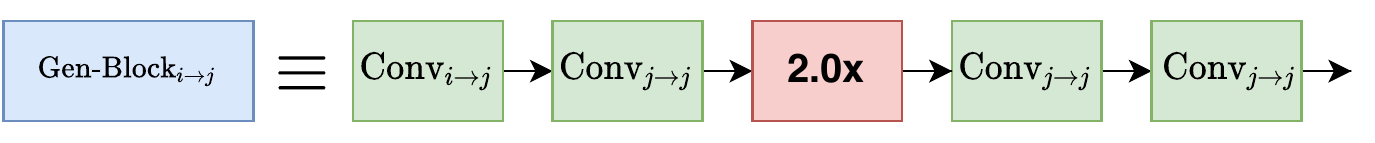}}
    \hspace{0.5cm}
    \subfloat[Critic block.]{\includegraphics[width=0.45\linewidth]{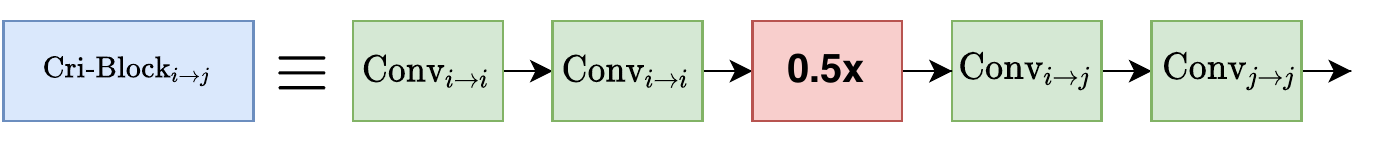}}
    \caption{Generator (a), critic (b), generator block (c), and critic block (d)  architectures. Subscript $i \to j$ denotes the feature dimension of the input ($i$) and output ($j$). Convs in the generator block are followed by batch norm and Leaky ReLU; Convs in the critic block are followed by Leaky ReLU.}
    \label{fig:model_arch}
\end{figure}

Our model is largely based off the BigGan \cite{brock2019large} architecture, following a similar channel progression and utilizing global sum pooling, although we replace 2D convolutions with 3D and remove the residual connections.
Our full generator, shown in Fig. \ref{fig:model_arch}a, consists of 20 convolutional layers organized into five sequential blocks, with block structure shown in \ref{fig:model_arch}c. It is built and trained in a block-wise progressive fashion \cite{DBLP:journals/corr/abs-1710-10196}. Block $i$ produces samples in $\mathbb{R}^{T \times H_i \times W_i}$, where $T$, $H_i$, and $W_i$ denote the number of timesteps, height, and width, respectively. The first block takes as input a 256-dimensional noise vector and produces output in $\mathbb{R}^{32 \times 4 \times 8}$ output. Note that even the first block's output is of the full temporal resolution, $T=32$, which initial experiments suggested performed better than progressively growing in the time dimension. The next four blocks each double the spatial resolution, yielding an overall generator output in $\mathbb{R}^{32 \times 64 \times 128}$. Every block halves the feature dimension.
The critic's architecture (Fig. \ref{fig:model_arch}b and d) largely mirrors the generator's: it consists of five sequential blocks, each of which halves the spatial dimensions of the data and doubles the feature dimension. Batch normalization is not used in the critic, following the findings of \cite{DBLP:journals/corr/GulrajaniAADC17}.
We use a block-wise progressive training process, the details of which can be found in Appendix~\ref{sec:progressive}. Hyperparameter details can be found in Appendix~\ref{sec:tuning}.

\begin{figure}
    \centering
    \includegraphics[width=0.90\textwidth]{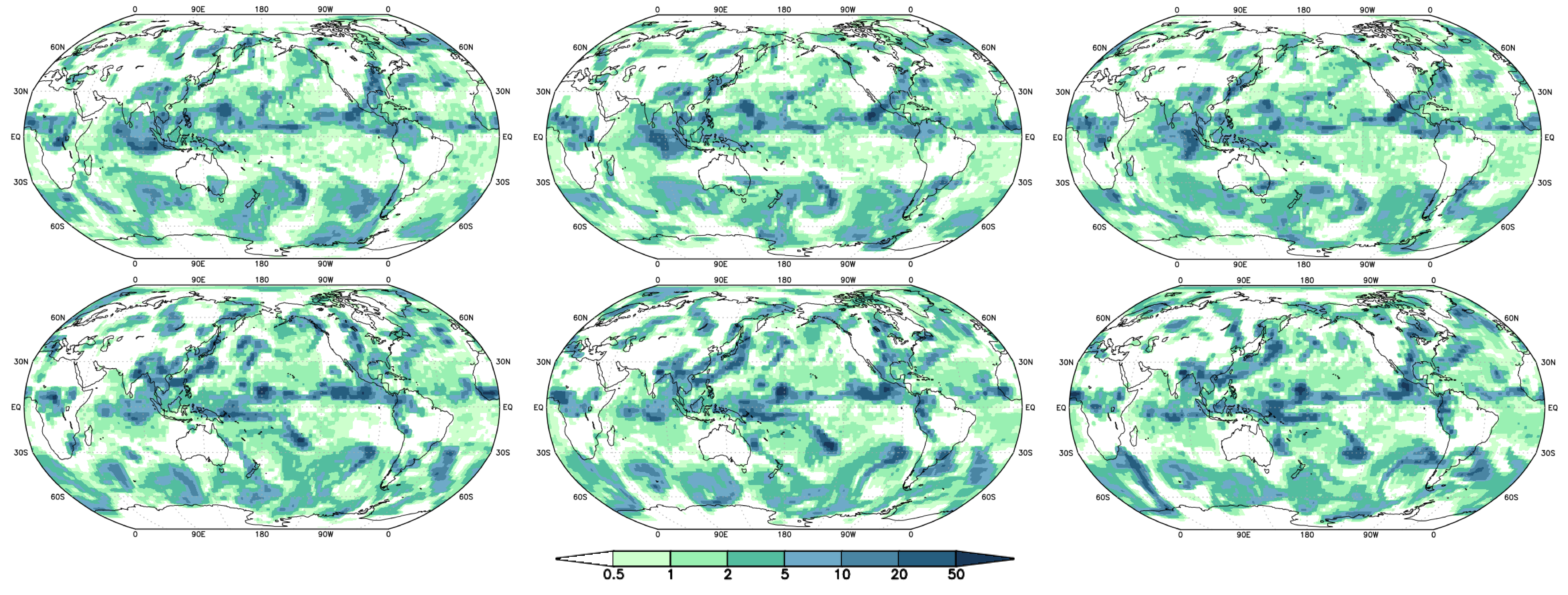}
    \caption{Generated samples. Top row: precipitation maps for three successive generated days. Bottom row: precipitation maps for three successive test days. Units: mm/day.}
    \label{fig:samples}
\end{figure}

\section{Experimental Setup}

\paragraph{Data}

We use daily output from the MIROC5 Earth System Model \cite{MIROC}, which has fully coupled atmosphere, ocean, land, and sea ice components at a $\sim$ 140 km horizontal resolution.  
Although we limit the current study to precipitation, future work will generalize to other variables (e.g., temperature and humidity).
We limit our study to the historical simulation (1850--2005), as described under the Coupled Model Intercomparison Project Phase 5 \cite{tayloretal2012}. To account for the highly skewed distribution of daily precipitation values (mm/day), we apply $\log(1 + x)$ normalization, but undo this normalization before computing performance metrics.
We randomly split the data, by 9-year-long chunks, into training (90\%), validation (5\%) and test (5\%) sets. We then split our data into two half-years, with different seasonal behaviors: Fall-Winter (SONDJF) and Spring-Summer (MAMJJA).
For Fall-Winter, this yields 131225, 8145, and 8145 days in training, validation and test, respectively.
For Spring-Summer, training, validation and test have 133400, 8280 and 8280 days, respectively.
One model is trained on the Fall-Winter data, another is trained independently on the Spring-Summer data; we refer to these models as ``loosely conditioned'' (on the season), as opposed to a hypothetical model more tightly conditioned (e.g., on a target amount of global precipitation).
Fig. \ref{fig:samples} contains generated and test set Fall-Winter samples.

\paragraph{Performance Metrics}

We use KL divergence to measure dissimilarity between pairs of empirical distributions (e.g. those induced by samples from a generator and those by samples from the test set). Our proposed approach is to take each sample in the first set, flatten it to a vector in $\mathbb{R}^{THW}$, and fit a full-covariance multivariate normal (MVN) distribution to the set of vectors. This process is repeated for the second set, and then we compute the KL divergence between the two distributions. However, estimating the $TWH \times TWH$ covariance matrix is impractical. Instead, we break each sample up into subtensors and fit MVN distributions over these smaller vectors. 
We consider three ways to break the overall tensor up into smaller tensors.

\begin{enumerate}
    \item {\bf Spatial KL}. 
    We first break each sample up by time, yielding $T$ spatial maps in $\mathbb{R}^{H \times W}$. If $H>32$ and $W>64$, each map is further broken into $(\frac{H}{32})(\frac{W}{64})$ maps, each in $\mathbb{R}^{32 \times 64}$. 
    \item {\bf Temporal KL}. 
    We break sample into $HW$ temporal vectors, each in $\mathbb{R}^T$.
    \item {\bf Spatiotemporal KL}. 
    Here we produce small subtensors with both time and space dimensions.  Specifically, we break each sample into  $(\frac{H}{4})(\frac{W}{8})$ subtensors, each in $\mathbb{R}^{T \times 4 \times 8}$. 
\end{enumerate}

\paragraph{Evaluation Process}
Recall that our goal is not to predict a particular outcome, but to produce a generator that defines a distribution as close to the distribution of the ESM as possible. To evaluate our ``loosely conditioned'' generators, we first generate two sets of $N$ samples:
one set using the generator trained on Fall-Winter data and 
the other the generator trained on the Spring-Summer data.
We compute the three KL divergences  between each of these sets of data and the two test sets (Fall-Winter and Spring-Summer). 
Because these values alone are not easy to interpret, we also compute and report the same performance metrics between the two half-year validation sets
and the test sets. Lastly, because validation and test are both produced by the same ESM, the ``performance'' of validation is effectively an upper bound on our performance. We therefore also consider a degraded version of the validation data, in which zero-mean Gaussian noise is added to each element of each sample. We pick $\sigma=0.024$ as the smallest standard deviation such that the degraded validation performance (according to the spatiotemporal KL) matches the generated data performance over both seasons.

\section{Results and Discussion}
\begin{figure}
    \centering
    \subfloat[Spatiotemporal]{\includegraphics[width=0.25\textwidth]{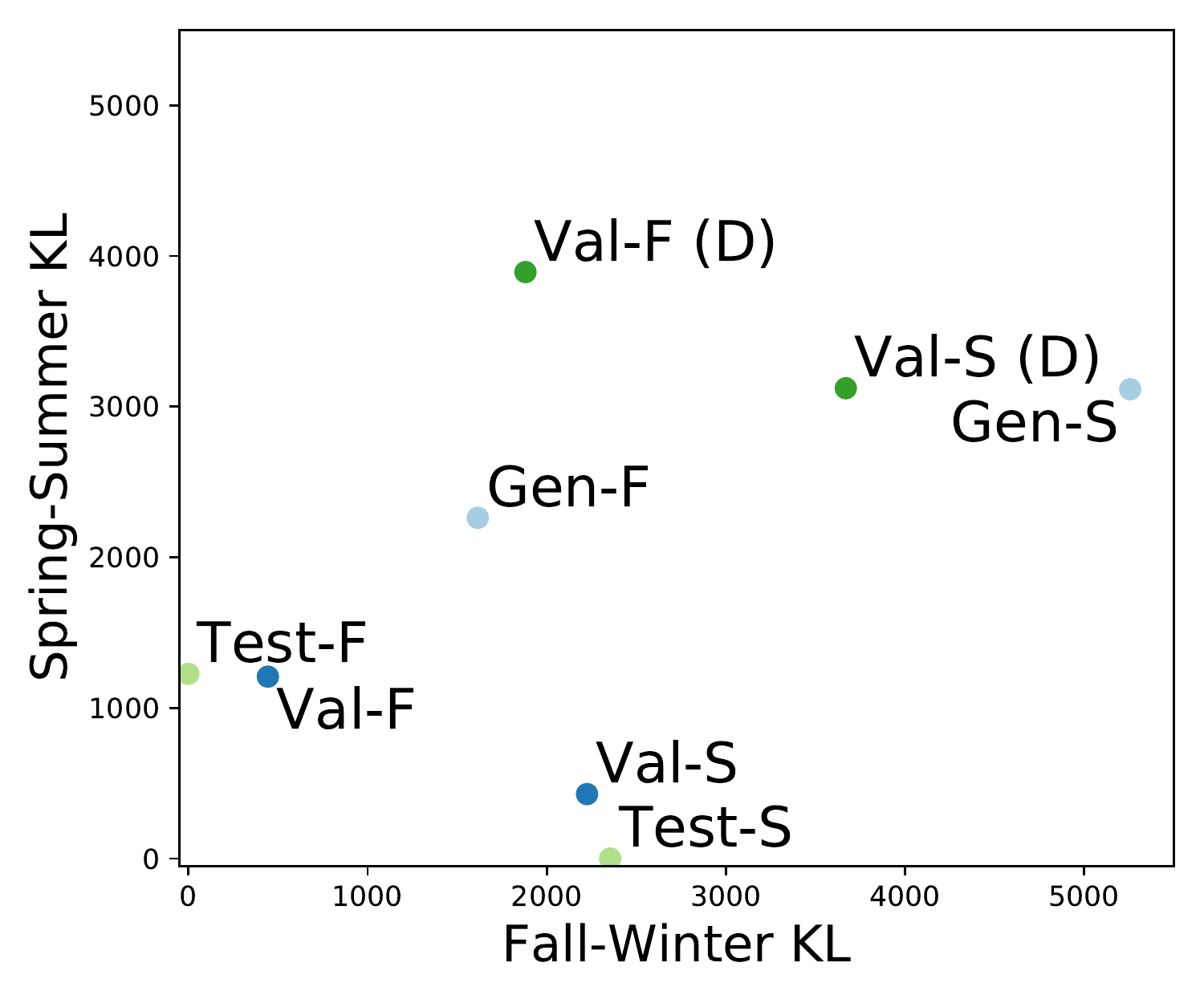}}
    \subfloat[Spatial]{\includegraphics[width=0.25\textwidth]{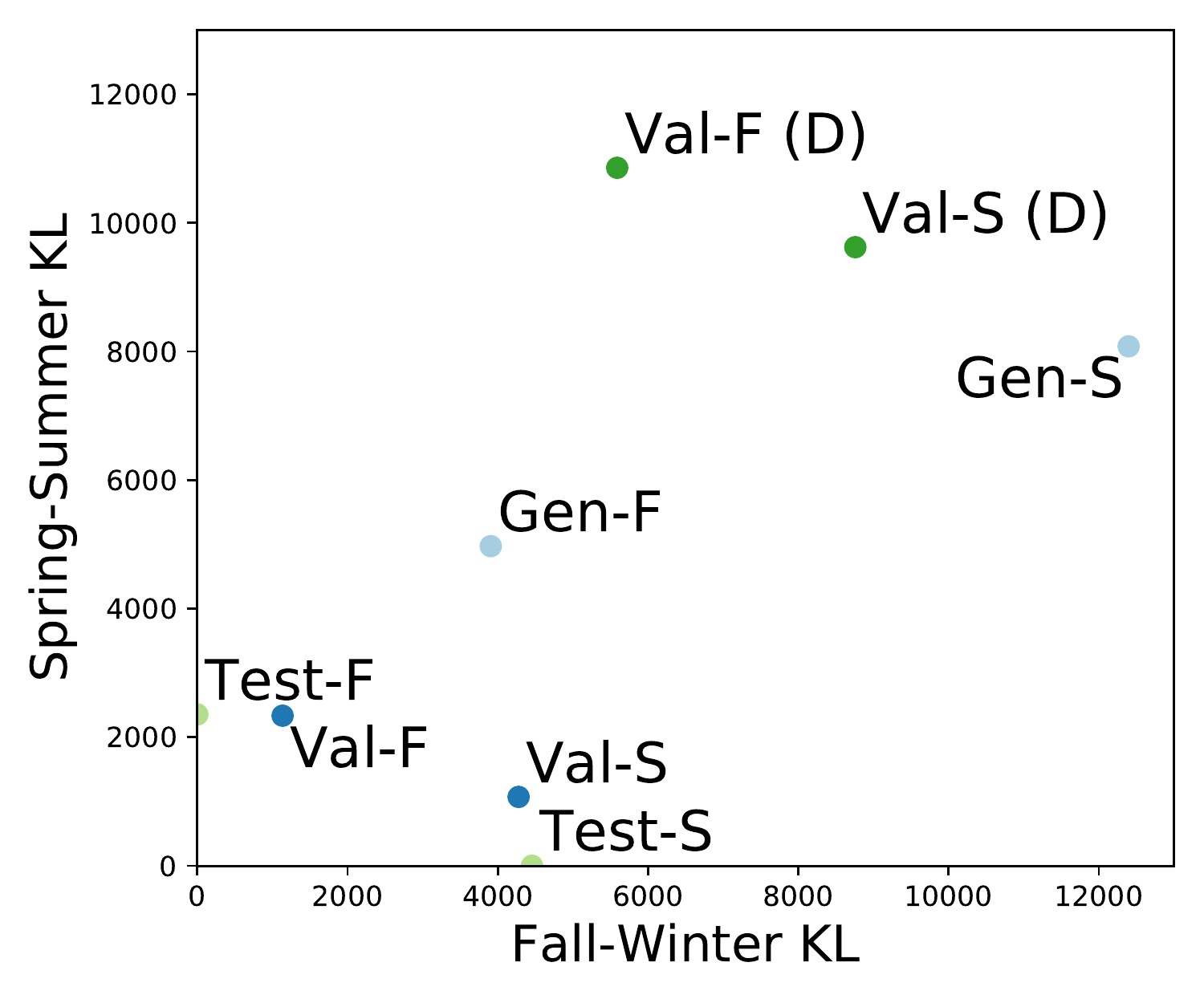}}
    \subfloat[Temporal]{\includegraphics[width=0.25\textwidth]{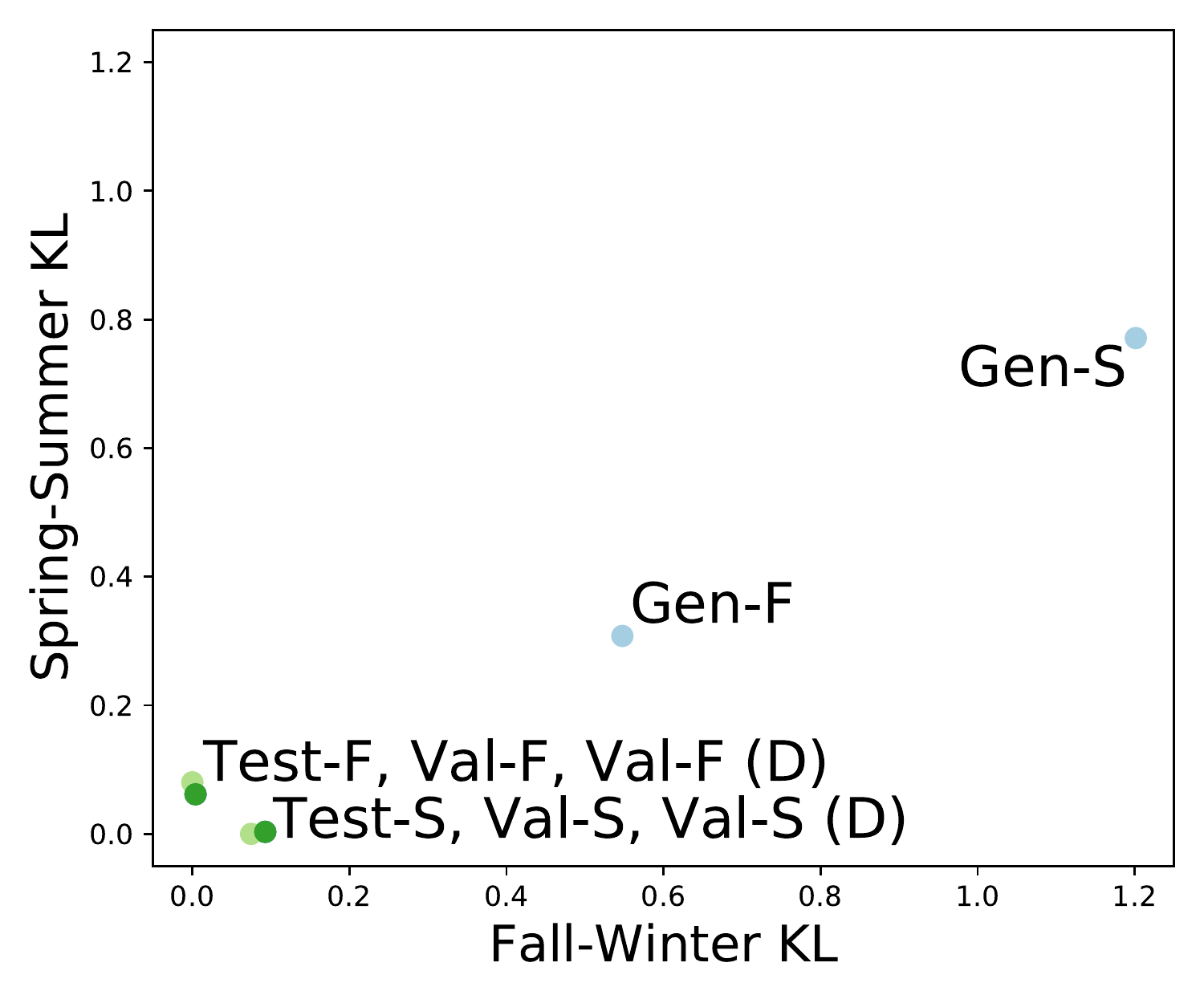}}
    \caption{KL divergences between Test and Validation, Validation plus noise $\sim {\cal N}(0,0.024)$ (Val (D)), Test and Generated data for Fall-Winter (F) and Spring Summer (S).}
    \label{fig:kl}
\end{figure}

Figure \ref{fig:kl} plots the three KL divergence metrics (a-c) for the Fall-Winter (F) and Spring-Summer (S) generated models. The x-axis coordinate represents the KL divergence between that empirical distribution and the Fall-Winter test set; likewise, the y-axis coordinate represents the KL divergence with the Spring-Summer test set. The ``(D)'' denotes the degraded version of validation (see above). We note that while the generated data does have higher spatial and spatiotemporal KL divergences than the clean validation data does, its performance is comparable to the validation data with only a very small amount of white noise added. By the temporal KL performance metric, the generated model is worse than the degraded validation data, which may be an result of our decision not to progressively grow the time dimension.
As expected, seasonality has an effect: the half-year data almost always has lower KL divergences with the matching half-year test set data.

\begin{figure}
    \centering
    \includegraphics[width=0.49\textwidth]{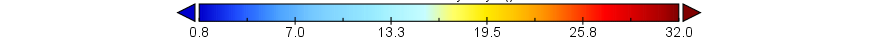}
    \includegraphics[width=0.49\textwidth]{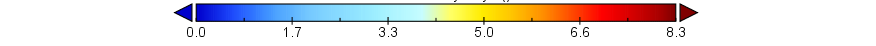}\\
    \vspace{-10pt}
    \subfloat[]{\includegraphics[width=0.24\textwidth]{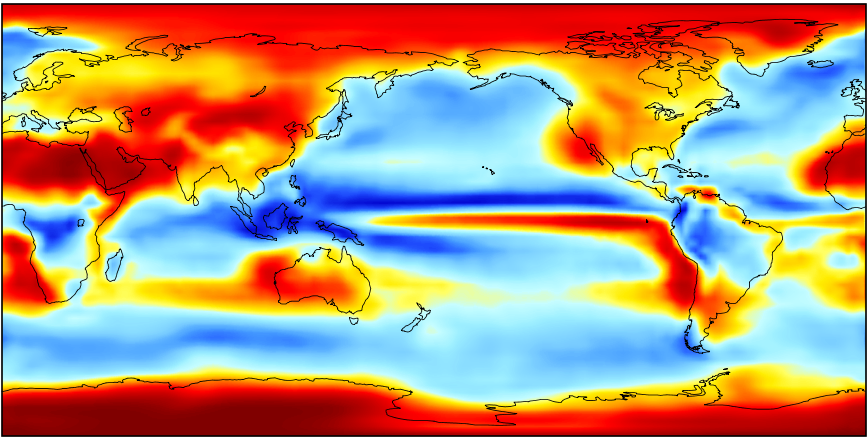}}
    \hspace{0.01cm}
    \subfloat[]{\includegraphics[width=0.24\textwidth]{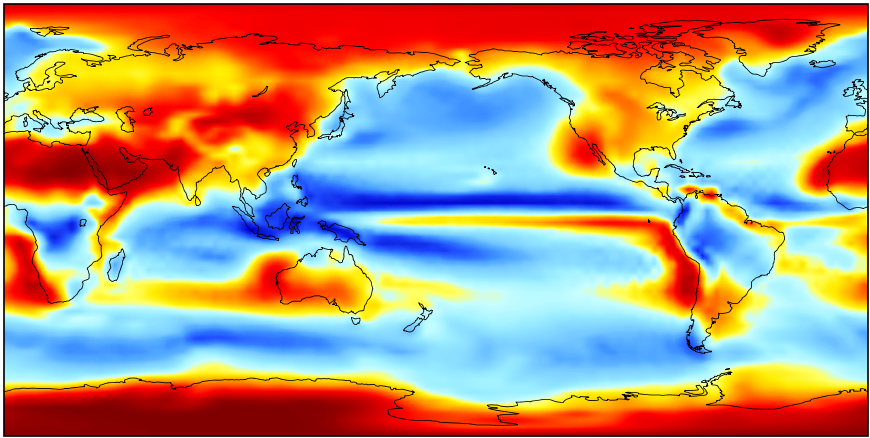}}
    \hspace{0.05cm}
    \subfloat[]{\includegraphics[width=0.24\textwidth]{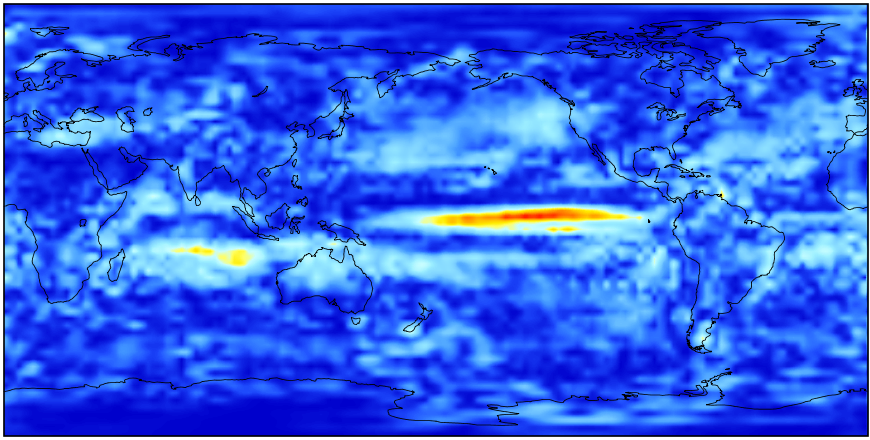}}
    \hspace{0.01cm}
    \subfloat[]{\includegraphics[width=0.24\textwidth]{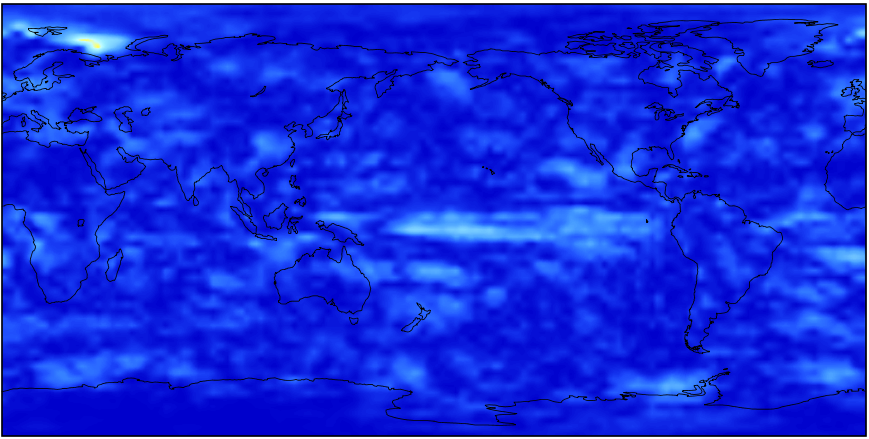}}
    \caption{Maps with mean number of dry days in the (a) generated and (b) test Fall-Winter data; (c) the absolute difference between a and b; (d) the absolute difference between test and validation.}
    \label{fig:maps_tot_dry}
\end{figure}

As an additional qualitative assessment of model performance, for each spatial location, we compute two statistics of the sets of samples: (i) the average number of dry days in each $T=32$ sample, and (ii) the average length of the longest dry spell in each sample (longest consecutive number of dry days). 
Fig.~\ref{fig:maps_tot_dry}a plots the  average number of dry days in the Fall-Winter generator's samples, and Fig.~\ref{fig:maps_tot_dry}b  the  average number of dry days in the Fall-Winter test data. The absolute difference between the two is plotted in panel c, and, for reference, panel d shows    
the absolute difference between these quantities when comparing test and validation data. Although the generated data is not quite as well matched as the validation data is, we see that most locations are no more than 1-3 days off (out of $T=32$). There is a notable exception, where errors approach 8 days in the equatorial Pacific west of South America, which appears to be due to a tendency of the generated data to have a slightly wider and intense dry tongue than the GCM climatology. Appendix D presents analogous maps and reports similar behavior for the maximum dry spell length statistics. Future work will address this behavior, here we note that for most applications of emulated dry days the domain of interest will be land regions of the globe, where our method does not manifest any strikingly biased behavior.

\section{Conclusion and Future Work}

We have demonstrated the ability of a progressive GAN architecture to successfully emulate the spatio-temporal characteristics of sequences of daily precipitation as produced by an ESM. Once trained, it becomes computationally efficient to produce an arbitrary number of such sequences, akin to the output of a large initial condition ensemble.  According to a number of verification methods, the GAN appears to have satisfactorily replicated the ESM's spatiotemporal behavior. 

There are many ways we plan to extend the current work.
The current work trains our GAN on a single earth system model and experiment; adding support for conditioning the generator would enable modeling arbitrary transient scenarios, an important next step. 
Established emulators of annual and monthly quantities on the basis of arbitrary scenarios of future forcing \cite{Linketal2019} could be used here as the sources of conditional information. 
While the current model is locked to producing a sequence of 32 days, it would be useful to develop models capable of producing any number of consecutive days. 
One last natural extension would be to explore the joint and sequential generation of variables (e.g. generating temperature and humidity conditioned on precipitation).
These capabilities, once developed, would support a significant improvement in  integrated modeling of climate change impacts by enabling a rich representation of some of the most damaging hazards and an exploration of uncertainties in scenario, model and internal variability~\cite{hawkinsandsutton2009} ``on the fly,'' substituting for expensive, time consuming and necessarily constraining computational resources needed to run climate models. 

\begin{ack}
This research was supported by the U.S. Department of Energy, Office of Science, as part of research in MultiSector Dynamics, Earth and Environmental System Modeling Program. The authors also thank the NVIDIA corporation for the donation of GPUs used in this work.
\end{ack}

\bibliographystyle{IEEEtran}
\bibliography{ref}

\appendix

\section{Progressive Training} \label{sec:progressive}

We employ progressive training \cite{DBLP:journals/corr/abs-1710-10196}. 
Specifically, we first train the first block of the generator and critic, which learn to produce and evaluate, respectively, samples with low spatial resolution.
After the first block has converged, we train the first two blocks of the generator and critic. This process continues until all blocks have been trained.

\begin{figure}
    \centering
    \subfloat[Fading of the generator model. The generator produces its low resolution sample which is spatially upscaled and weighed with the output of the new block]{\includegraphics[width=0.45\linewidth]{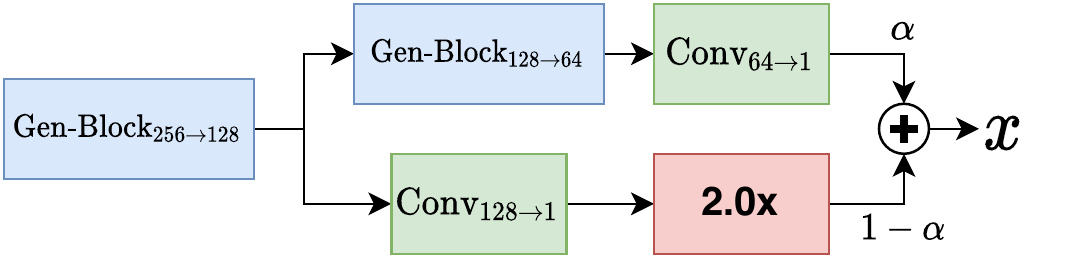}}
    \hspace{0.5cm}
    \subfloat[Fading of the critic model. The critic slowly begins using the features from the new block.]{\includegraphics[width=0.45\linewidth]{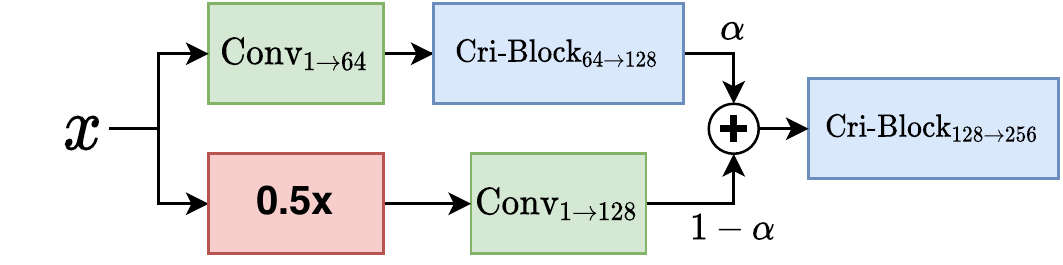}}
    \caption{Fading in additional layers during progressive training. $\alpha \rightarrow 1$ during the fading period.}
    \label{fig:fading}
\end{figure}

Importantly, when a new block is added, the training undergoes a 12,800 update {\it fading period}. 
During this fading period, the generated samples are a linear interpolation of the higher resolution but less trained output of the new block and the lower resolution but better trained output of the previous block. An equivalent process happens for the critic. The interpolation process is illustrated in Fig. \ref{fig:fading}.
The interpolation coefficient, $\alpha$, linearly transitions from 0 (start of fading period) to 1 (end of fading period).
After this fading period, $\alpha$ is effectively clamped to 1, and the block continues to train for up to 64,000 additional updates, until convergence. 
The fading period reduces the risk of the new block's randomly initialized weights disrupting training. Progressive training means that the early training is limited to computational cheap blocks, accelerating the training process.

\section{Hyperparameter Tuning} \label{sec:tuning}

We train each block with early stopping based upon the spatiotemporal KL divergence with the validation set.
We utilize a batch size of 32 and the AdamW optimizer. Our generator and critic learning rate is 8e-5, our grad penalty is 20.0, and for each iteration our critic is updated three times and the generator is updated one time. These parameters were found over a period of trial-and-error and hyperparameter searches utilizing Weights and Biases \cite{wandb} sweeps, which implements Bayesian Optimization \cite{Snoek12}. For the last block, we found that increasing the grad penalty to 40.0 greatly improved training stability.

\section{Empirical Cumulative Distributions}\label{sec:metrics}

\begin{figure}
    \centering
    \includegraphics[width=0.4\textwidth]{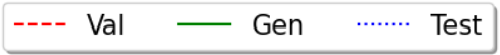}\\
    \vspace{-10pt}
    \subfloat[Denver]{\includegraphics[width=0.24\textwidth]{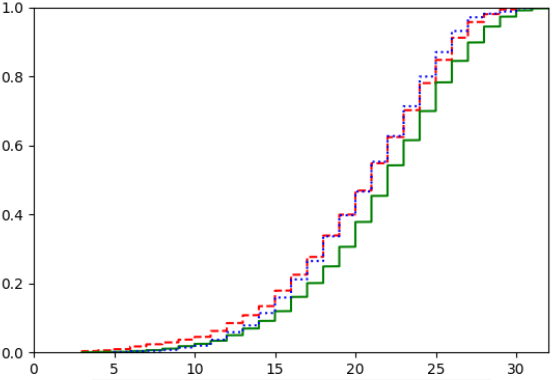}}
    \hspace{0.01cm}
    \subfloat[London]{\includegraphics[width=0.24\textwidth]{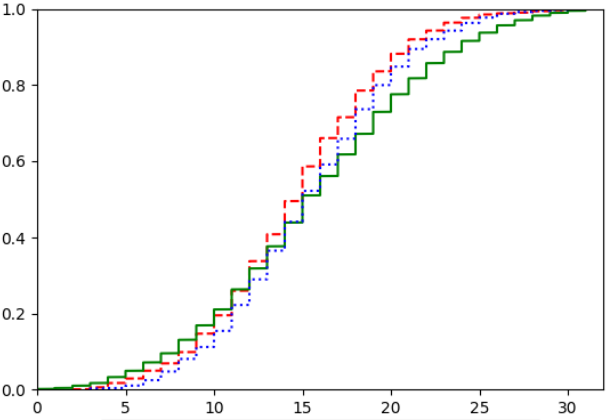}}
    \hspace{0.01cm}
    \subfloat[Denver]{\includegraphics[width=0.24\textwidth]{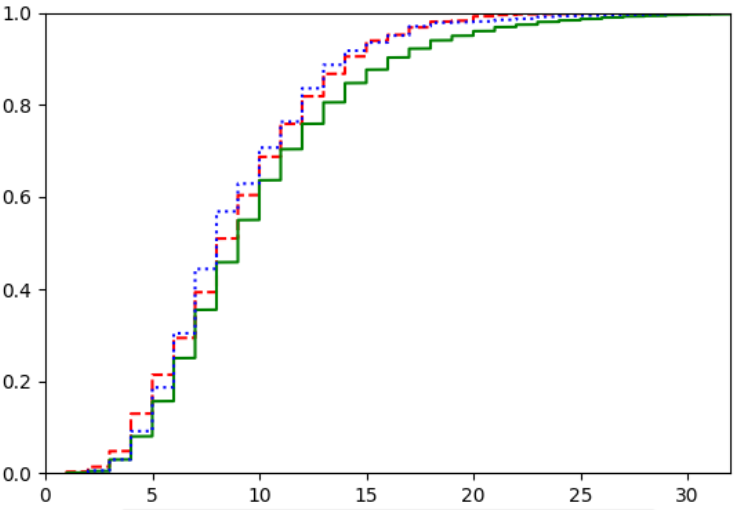}}
    \hspace{0.01cm}
    \subfloat[London]{\includegraphics[width=0.24\textwidth]{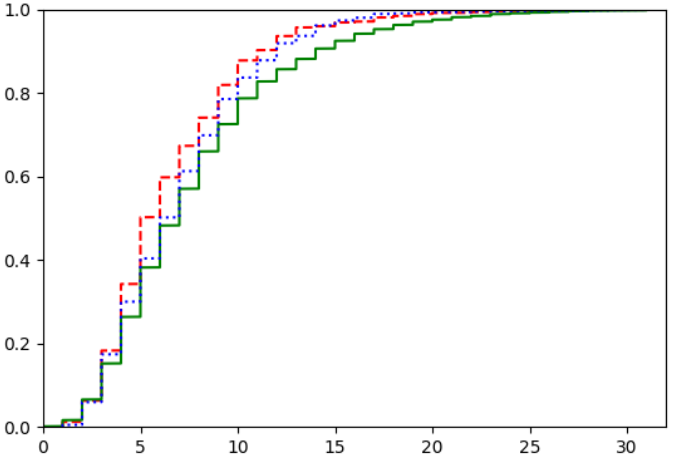}}\\
    \caption{Empirical cumulative distribution functions for the validation, generated and test data at two locations.  Number of dry days per sample in (a)-(b); length of longest dry spell in (c)-(d).}
    \label{fig:eCDFs}
\end{figure}

Continuing our analysis on performance, an additional metric compares the empirical distributions over points of interest in the test, validation and generated data. We choose two locations roughly corresponding to the coordinates of Denver, USA and London, UK. For the corresponding grid-points we extract time series and compute the total number of dry days (defined as before to be days with $<1mm$ precipitation) in each $T=32$ day sample. Figs. \ref{fig:eCDFs}a (Denver) and \ref{fig:eCDFs}b (London) plot the empirical cumulative distribution functions (eCDFs) of the number of dry days computed over $N=4096$ samples each from the generator, test and validation sets.  Figures \ref{fig:eCDFs}c and \ref{fig:eCDFs}d replicate this setup, except modeling the distribution over the longest dry spell in each sample.  

From the results we can assess a slight tendency of the generated samples to produce more dry days and longer dry spells than the samples from the climate model (both test and validation), but the difference between the eCDFs remains limited to very few days in all instances, and the shape of the eCDFs qualitatively similar. 

\section{Longest Dry Spell Analysis}\label{sec:metrics}

\begin{figure}
    \centering
    \includegraphics[width=0.49\textwidth]{0_to_32_legend.png}
    \includegraphics[width=0.49\textwidth]{drazic_diffs_legend.png}\\
    \vspace{-10pt}
    \subfloat[]{\includegraphics[width=0.24\textwidth]{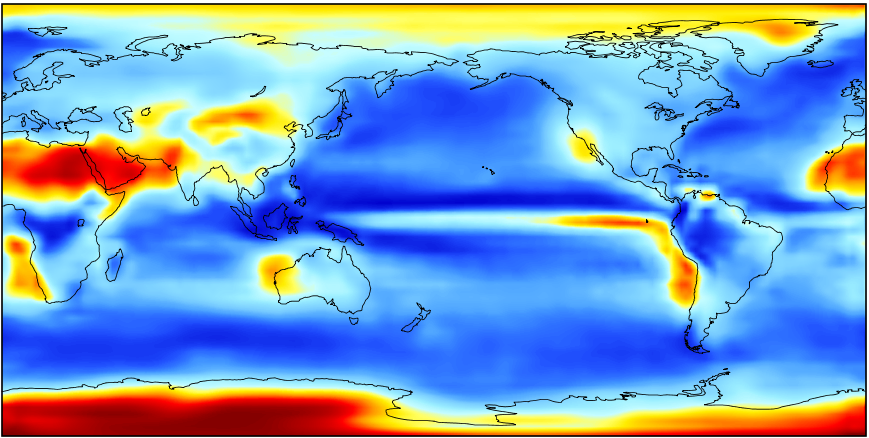}}
    \hspace{0.01cm}
    \subfloat[]{\includegraphics[width=0.24\textwidth]{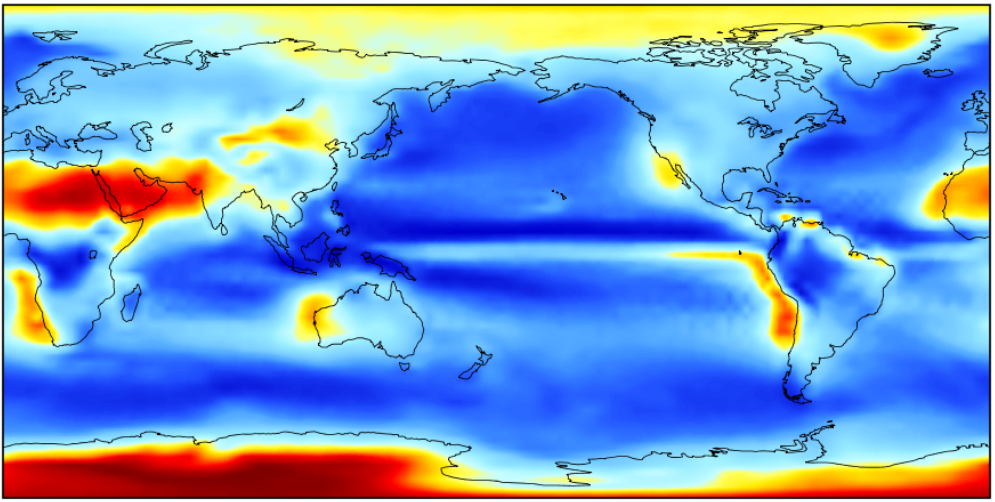}}
    \hspace{0.05cm}
    \subfloat[]{\includegraphics[width=0.24\textwidth]{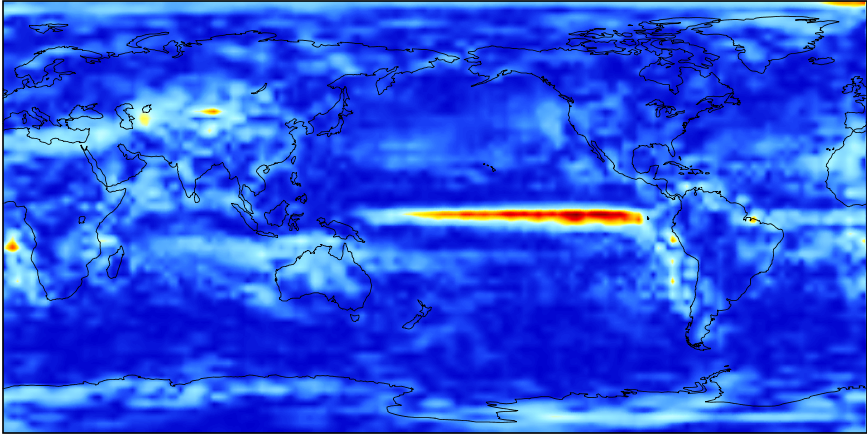}}
    \hspace{0.01cm}
    \subfloat[]{\includegraphics[width=0.24\textwidth]{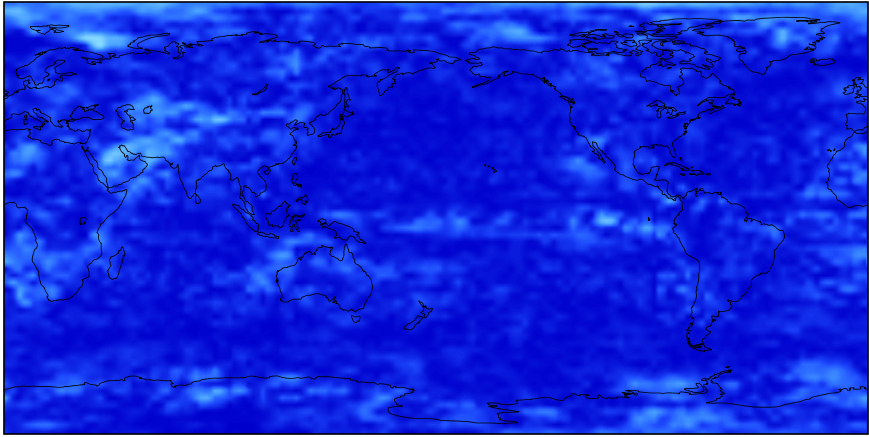}}
    \caption{Maps with mean longest dry spell length in the (a) generated and (b) test Fall-Winter data; (c) the absolute difference between a and b; (d) the absolute difference between test and validation.}
    \label{fig:maps_longest_dry_spell}
\end{figure}

Figure \ref{fig:maps_longest_dry_spell} plots maps of the average length of the longest dry spell from the Fall-Winter data, analogous to results presented in Fig. \ref{fig:maps_tot_dry}. The map of generated data averages is plotted in panel (a), whereas the map of test set averages is plotted in panel (b). Similar trends continue in panel (c), where the equatorial Pacific west of South America still provides the generator some difficulty.

\section{Additional Samples}

\begin{figure}
    \centering
    \includegraphics[width=0.45\textwidth]{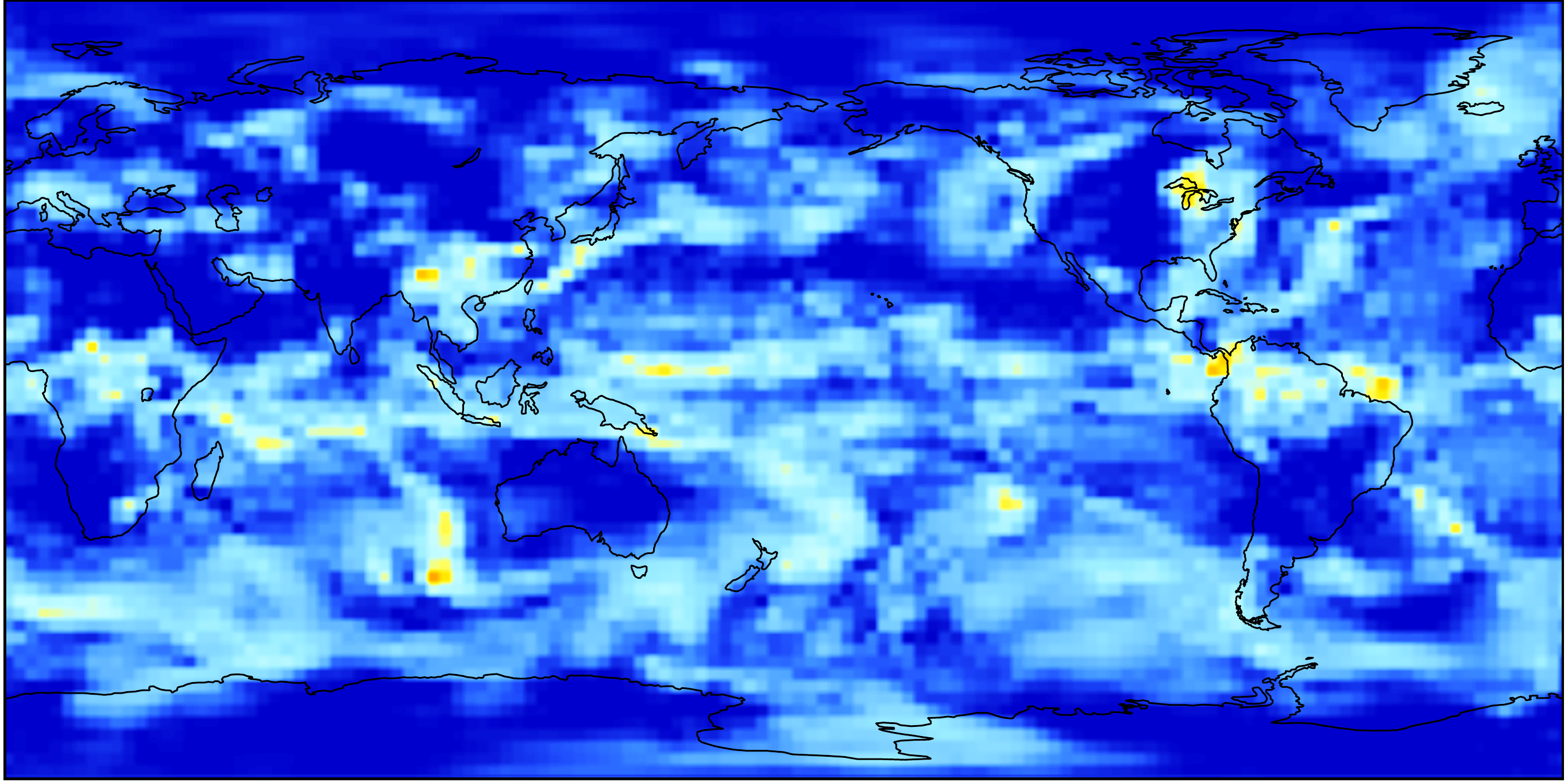}
    \includegraphics[width=0.45\textwidth]{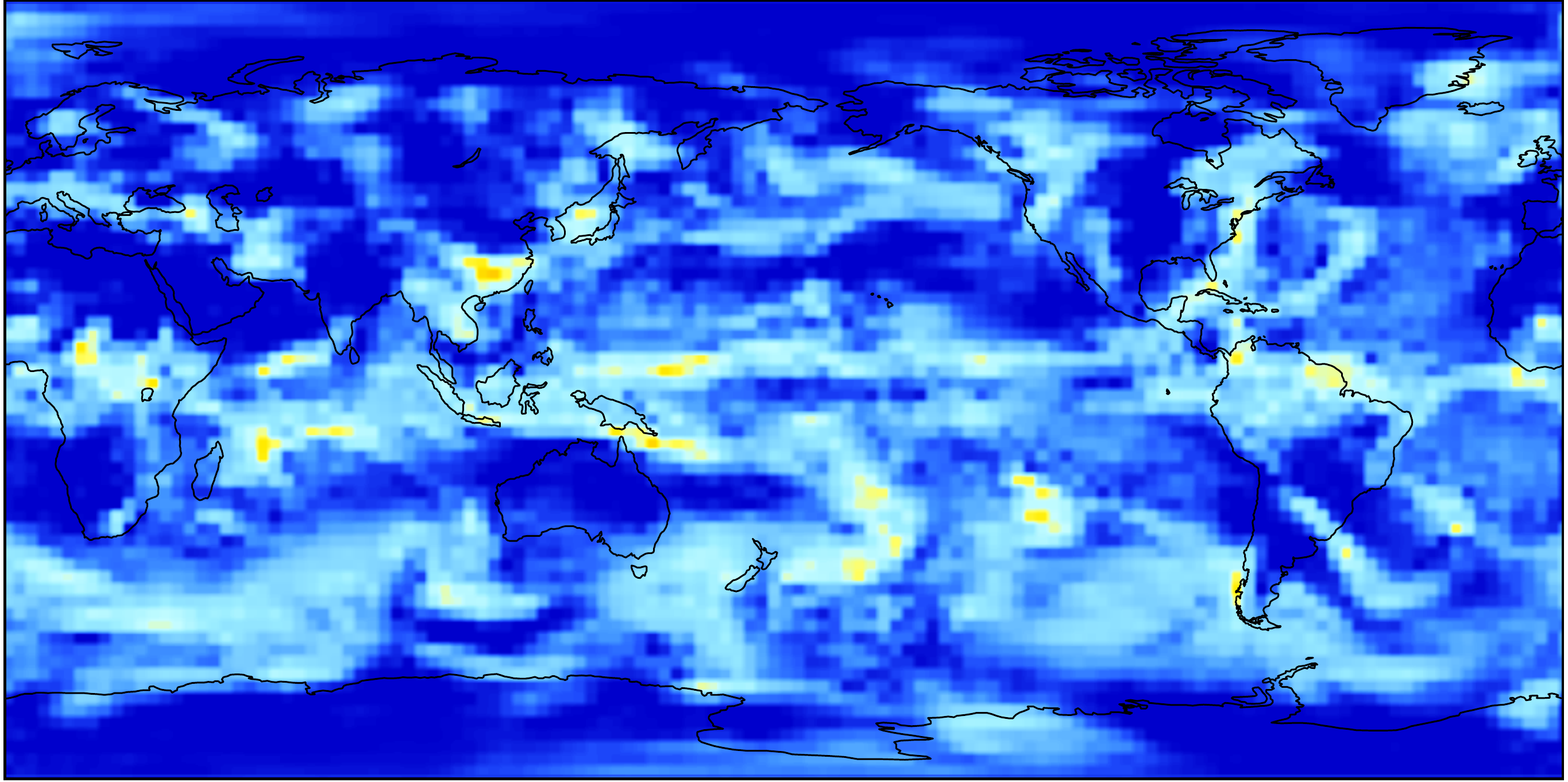}
    \includegraphics[width=0.45\textwidth]{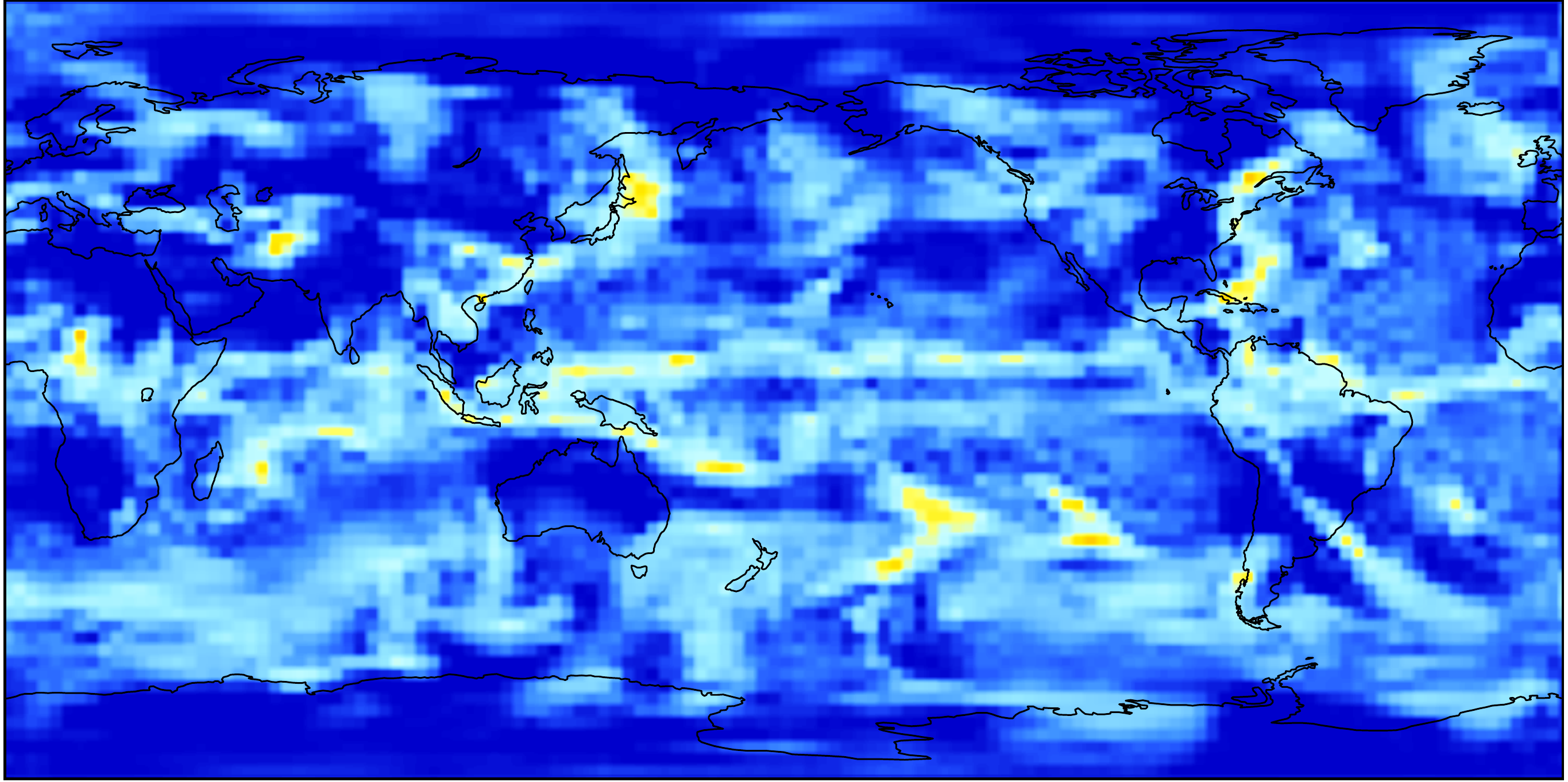}
    \includegraphics[width=0.45\textwidth]{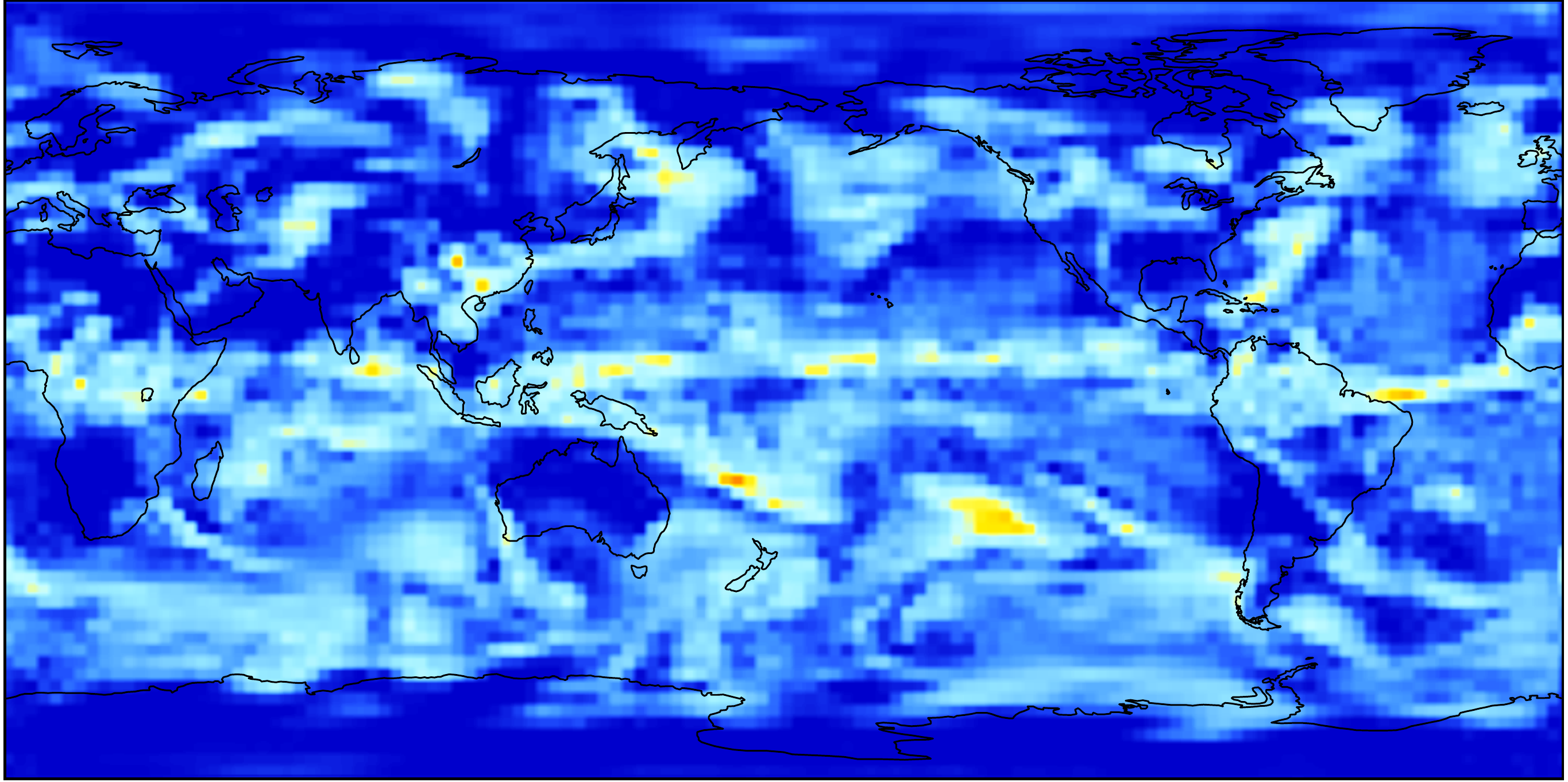}
    \caption{Precipitation maps for four successive generated days in spring and summer}
    \label{fig:gen_spring_0}
\end{figure}

\begin{figure}
    \centering
    \includegraphics[width=0.45\textwidth]{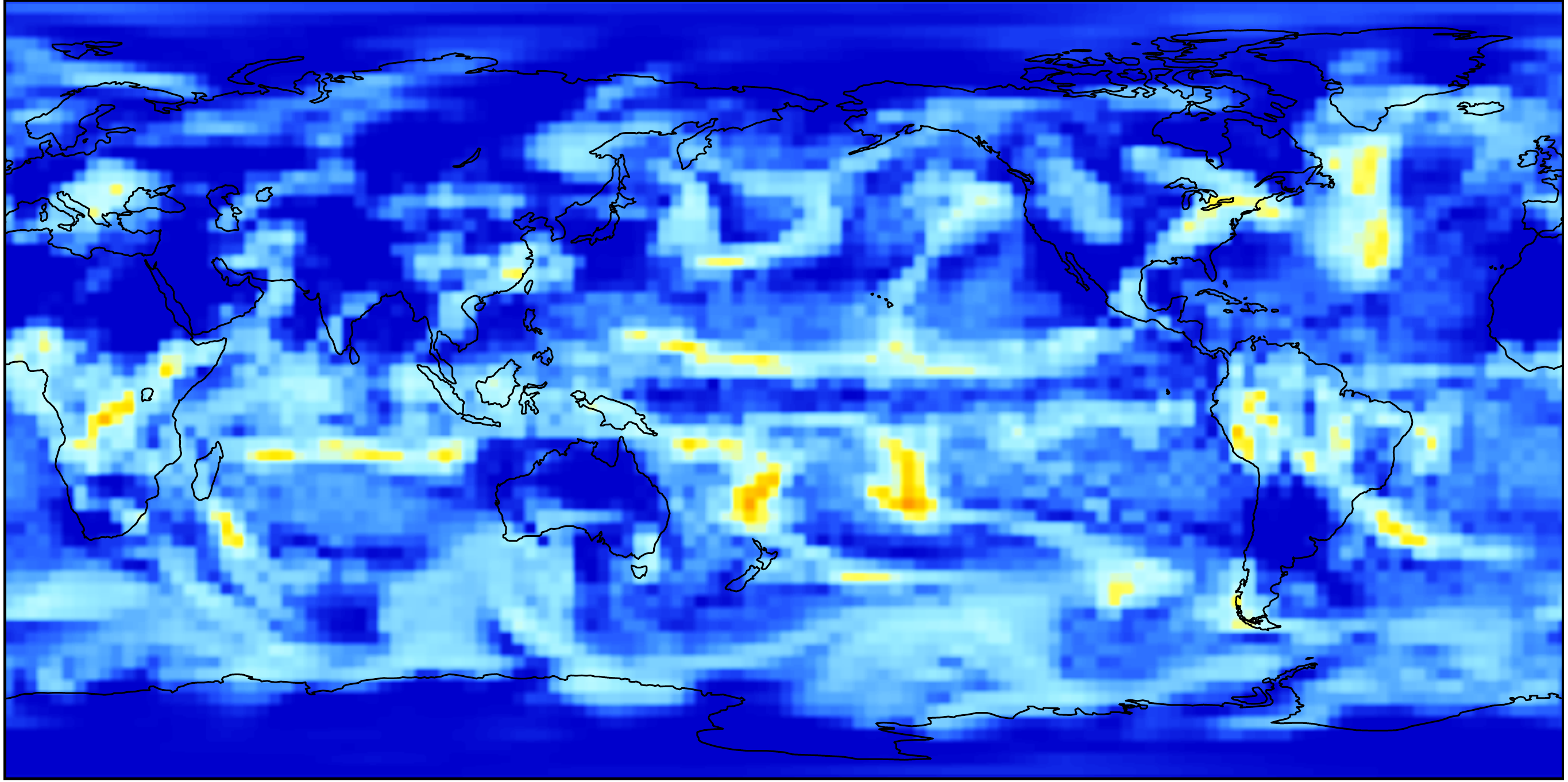}
    \includegraphics[width=0.45\textwidth]{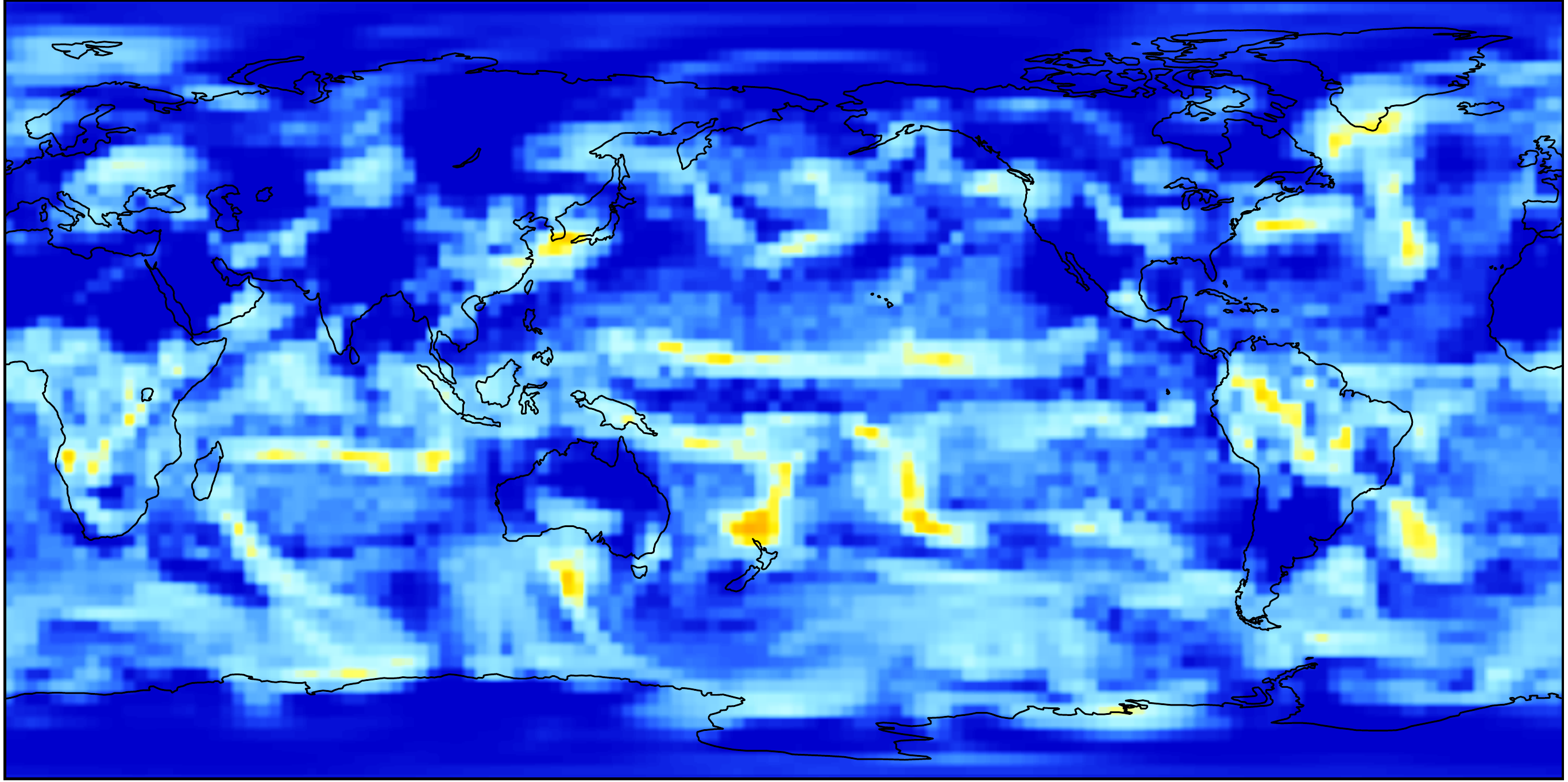}
    \includegraphics[width=0.45\textwidth]{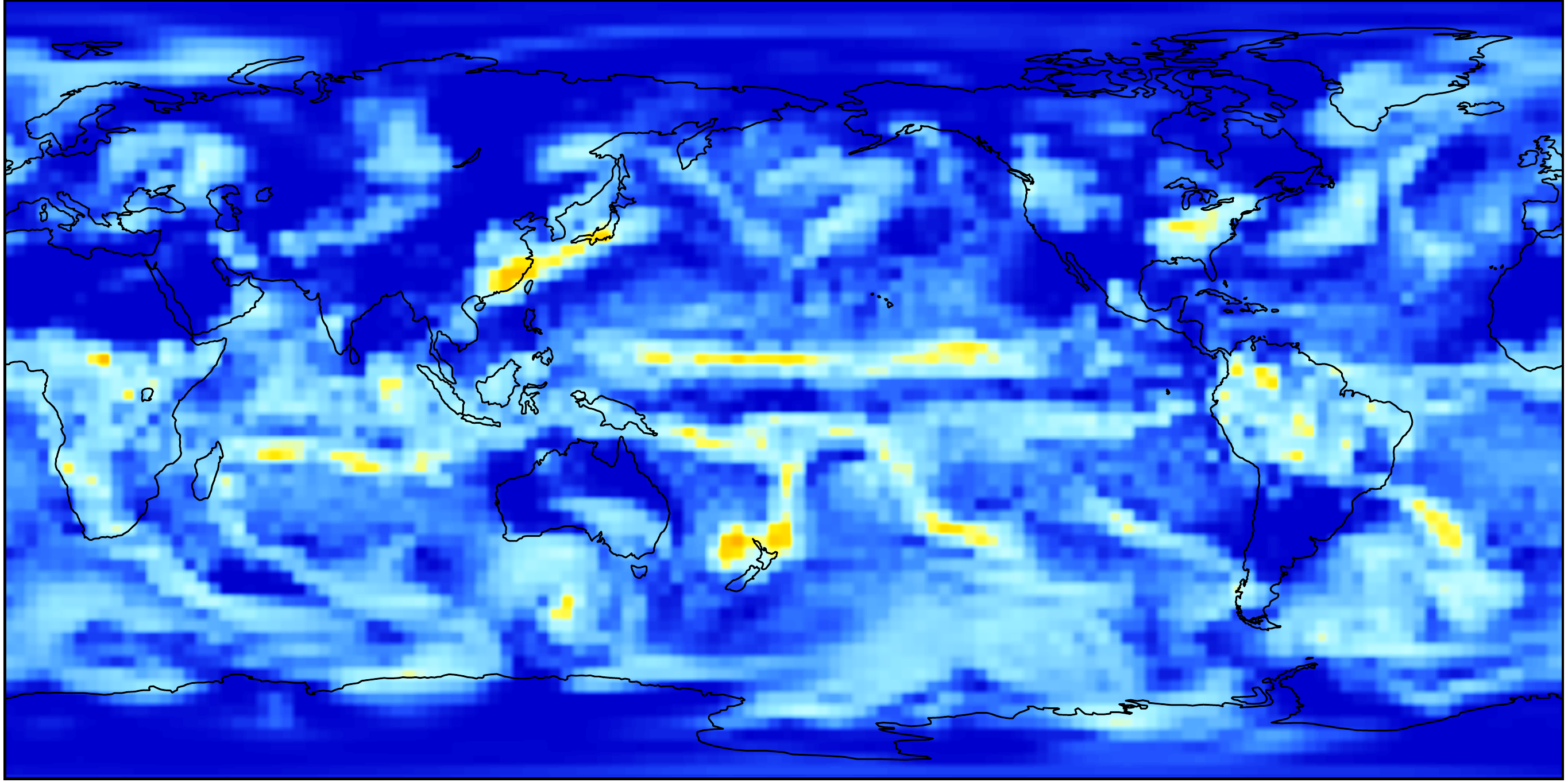}
    \includegraphics[width=0.45\textwidth]{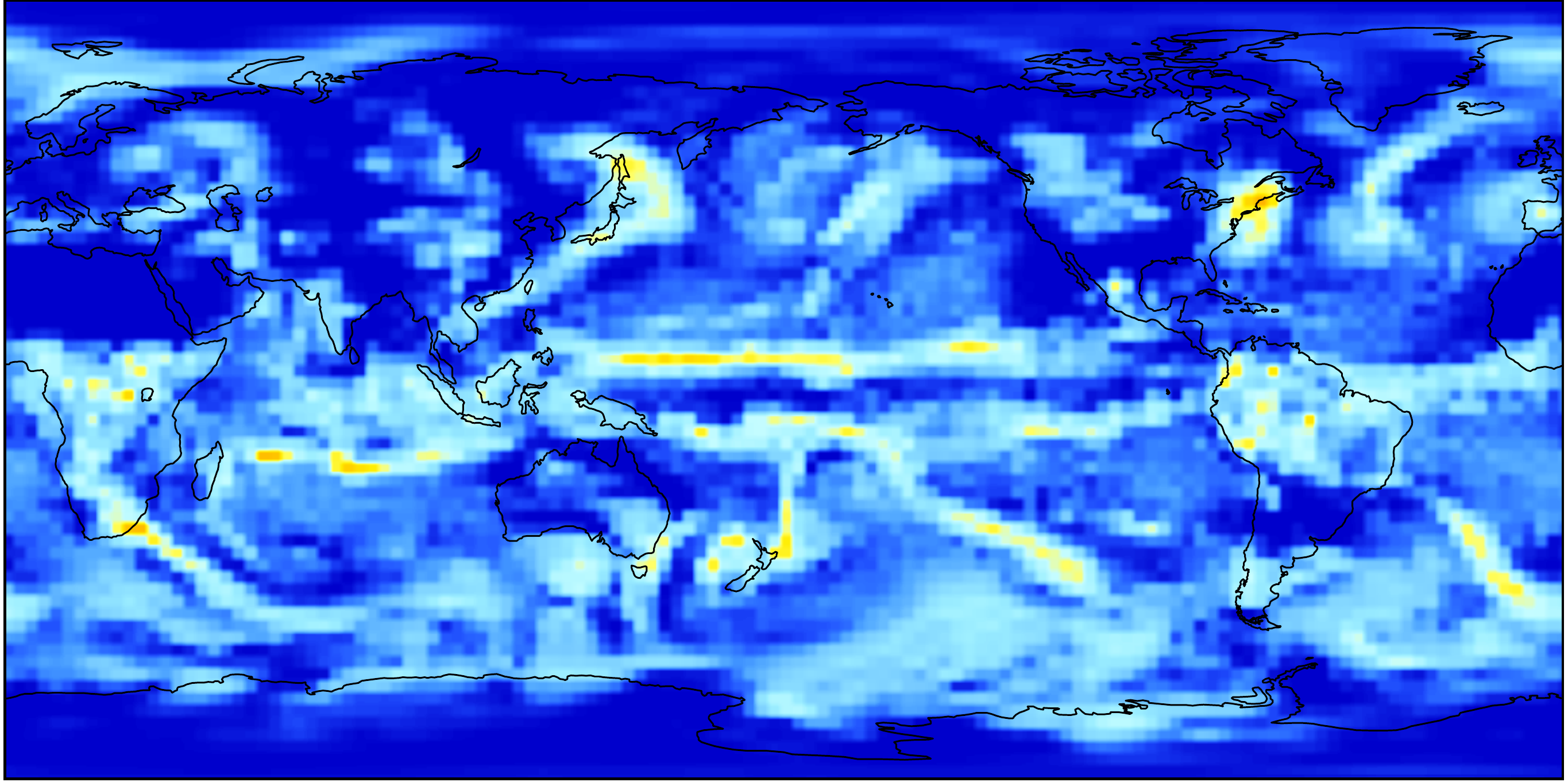}
    \caption{Precipitation maps for four successive test set days in spring and summer}
    \label{fig:real_spring_0}
\end{figure}

\begin{figure}
    \centering
    \includegraphics[width=0.45\textwidth]{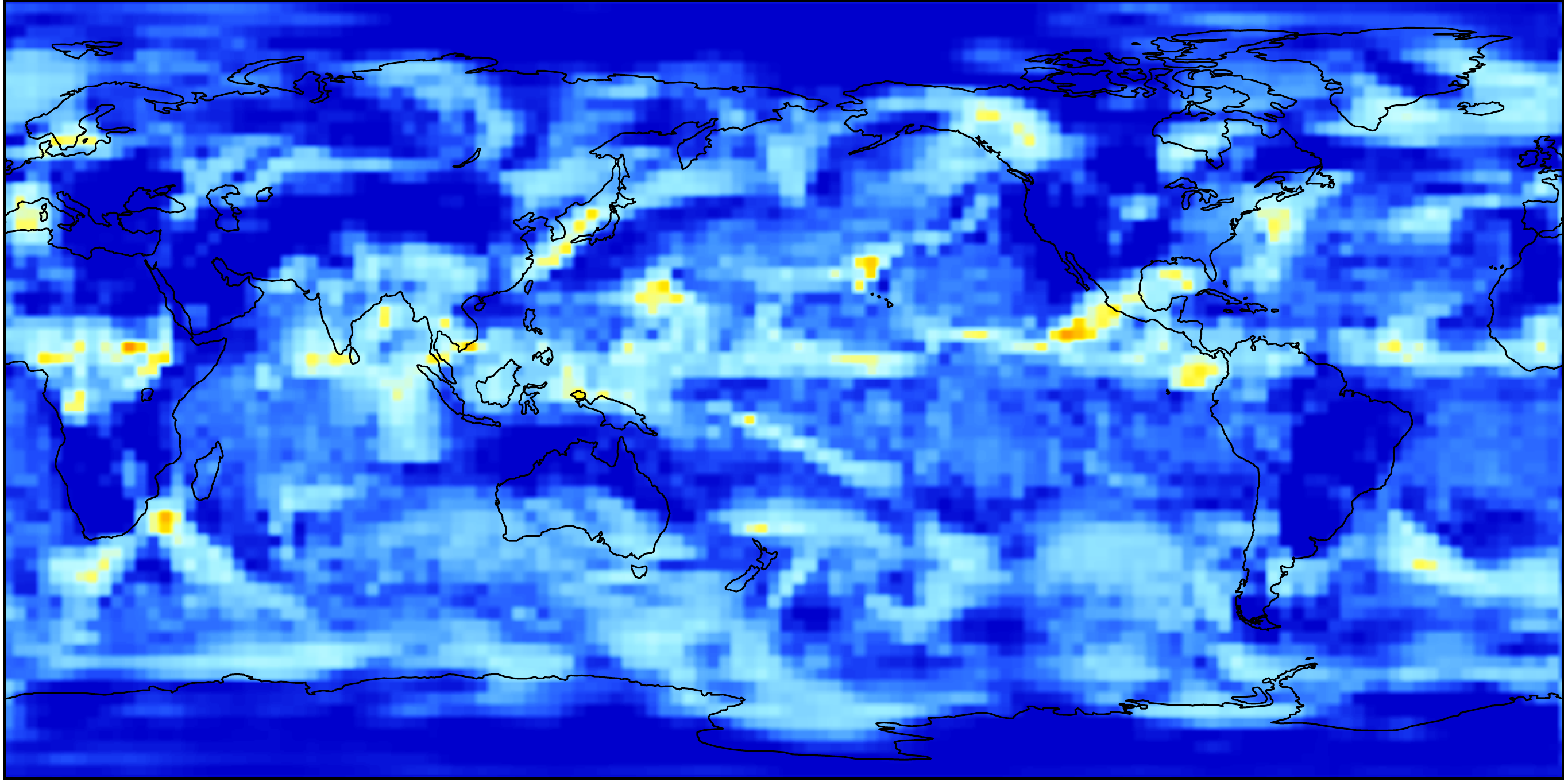}
    \includegraphics[width=0.45\textwidth]{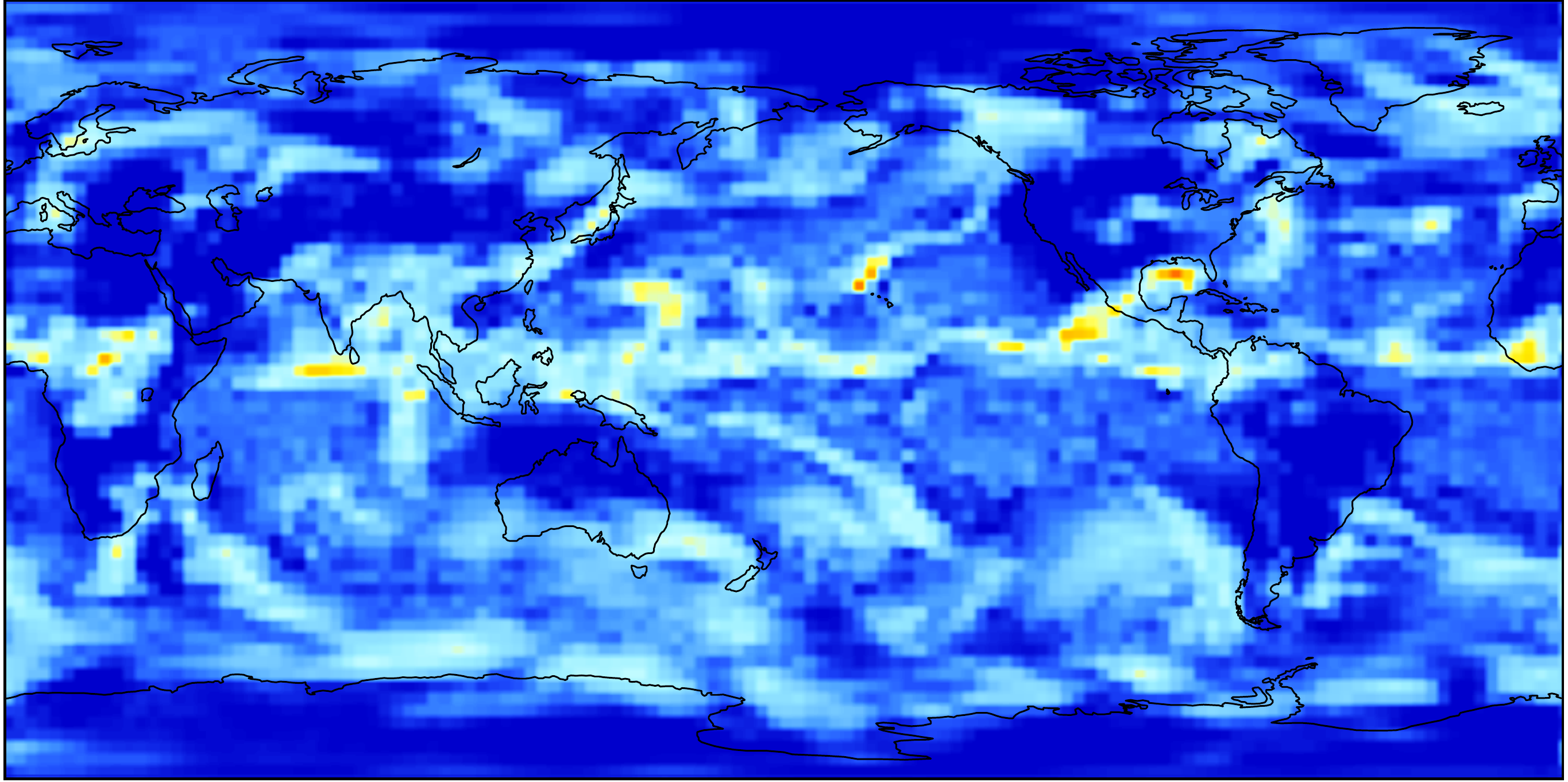}
    \includegraphics[width=0.45\textwidth]{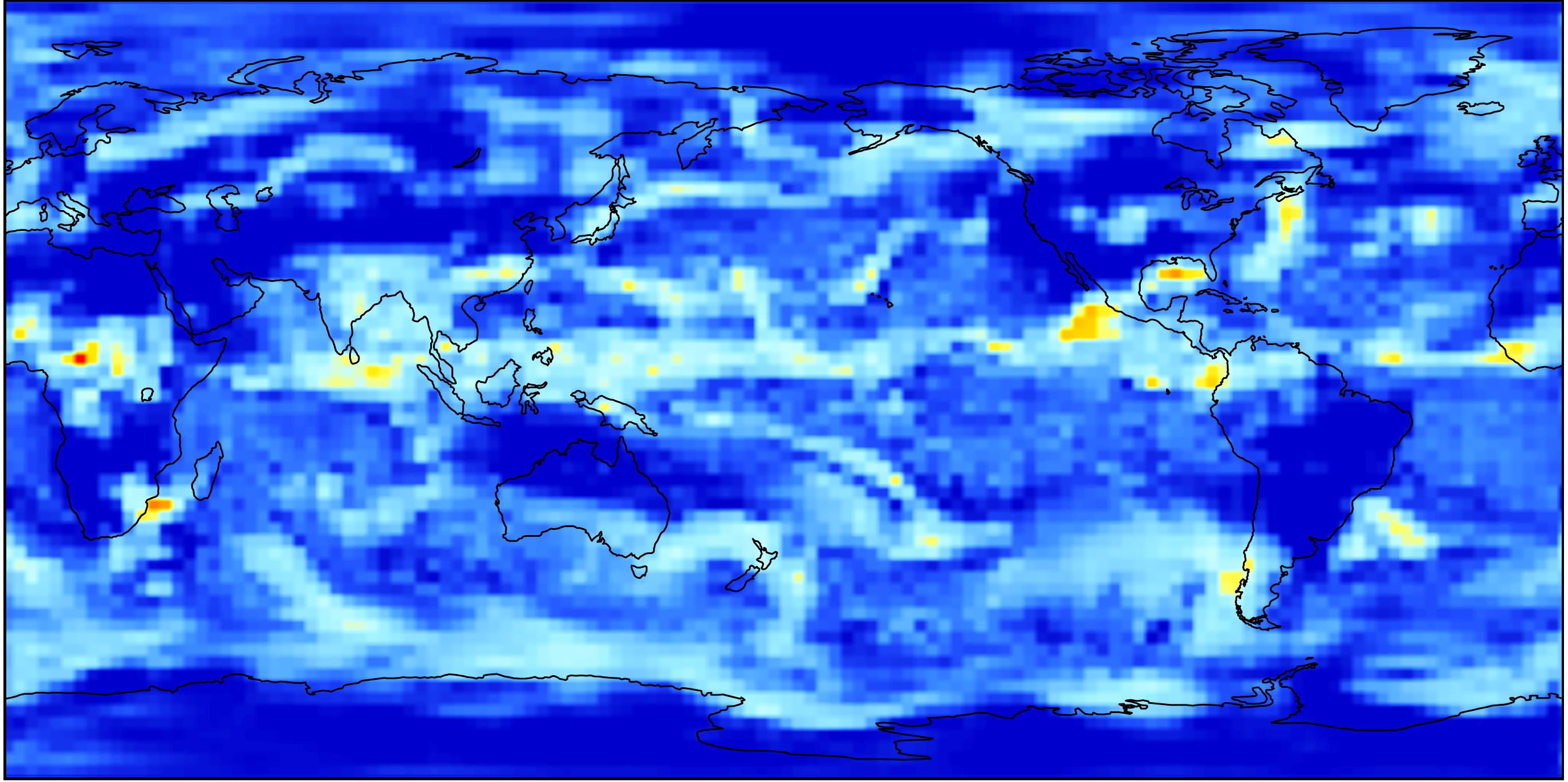}
    \includegraphics[width=0.45\textwidth]{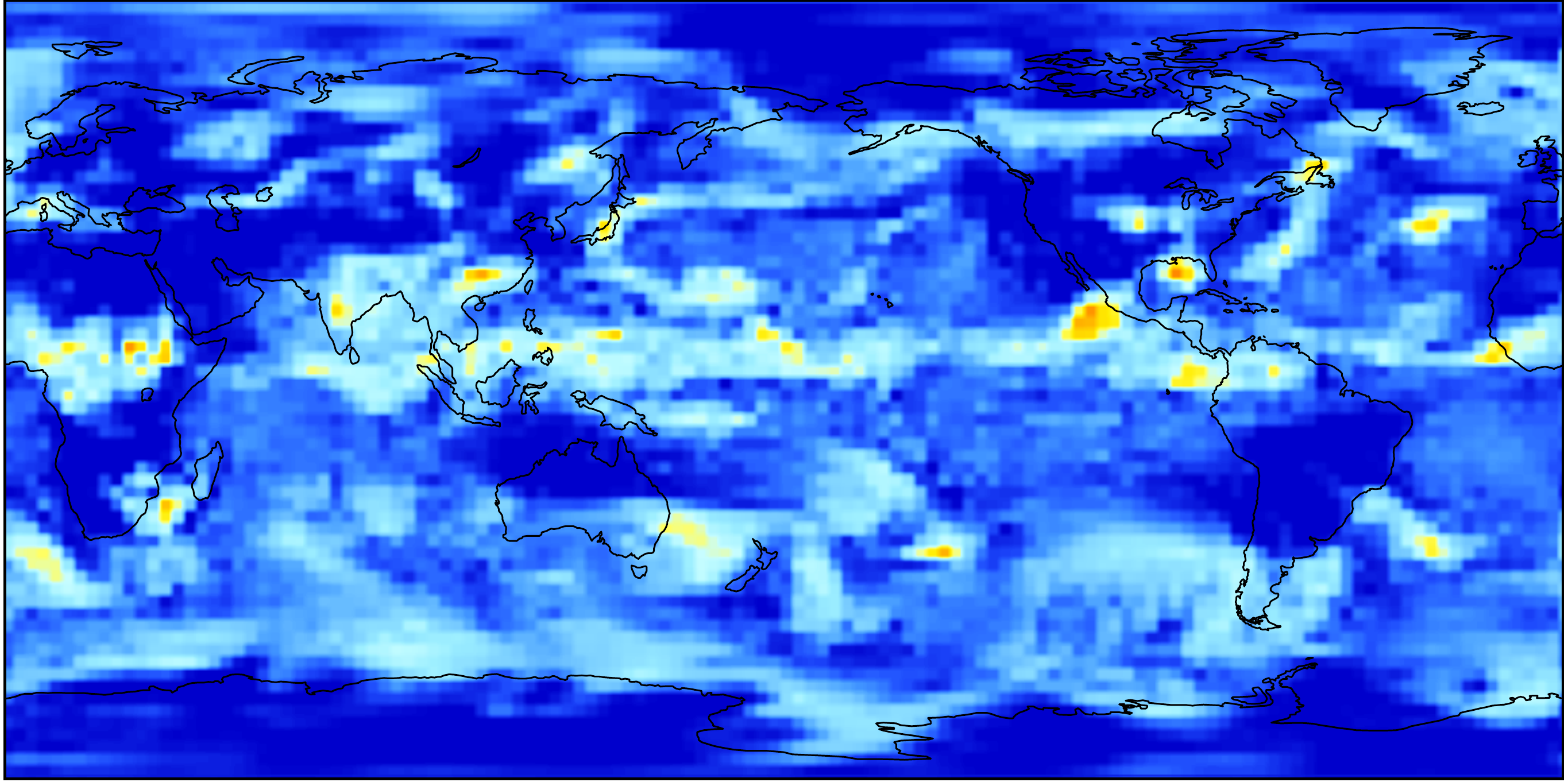}
    \caption{Precipitation maps for four successive generated days in fall and winter}
    \label{fig:gen_fall_0}
\end{figure}

\begin{figure}
    \centering
    \includegraphics[width=0.45\textwidth]{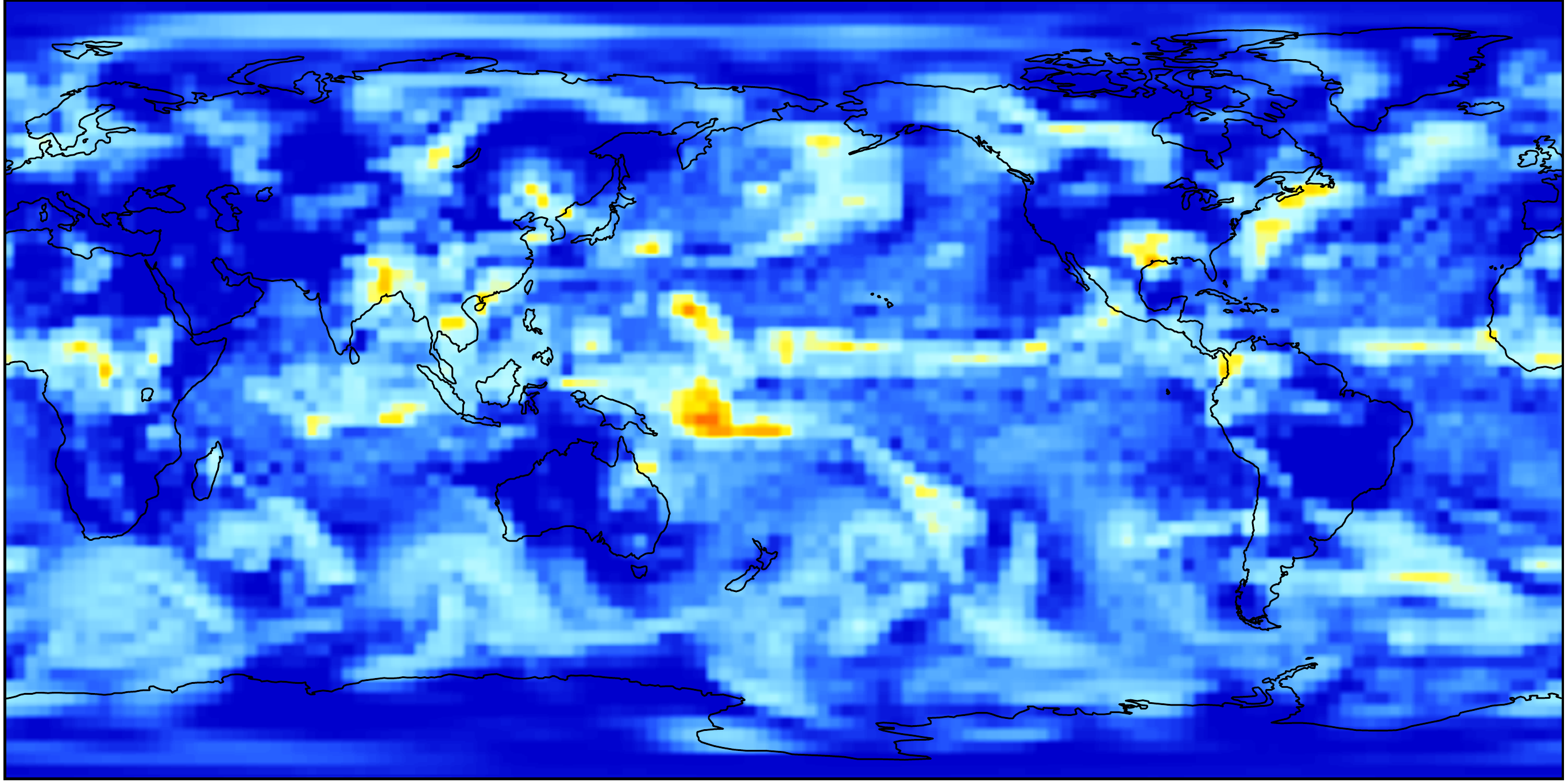}
    \includegraphics[width=0.45\textwidth]{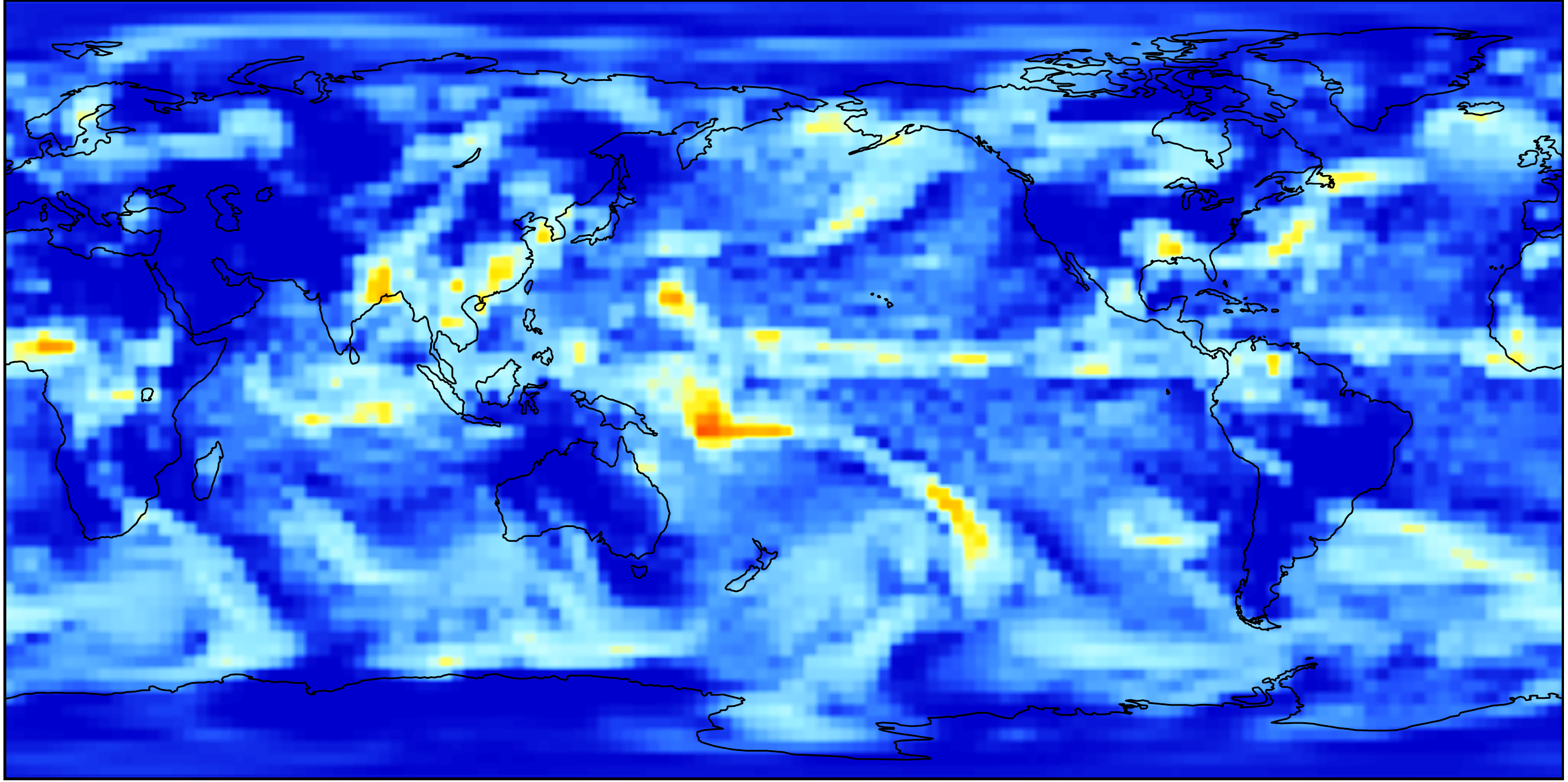}
    \includegraphics[width=0.45\textwidth]{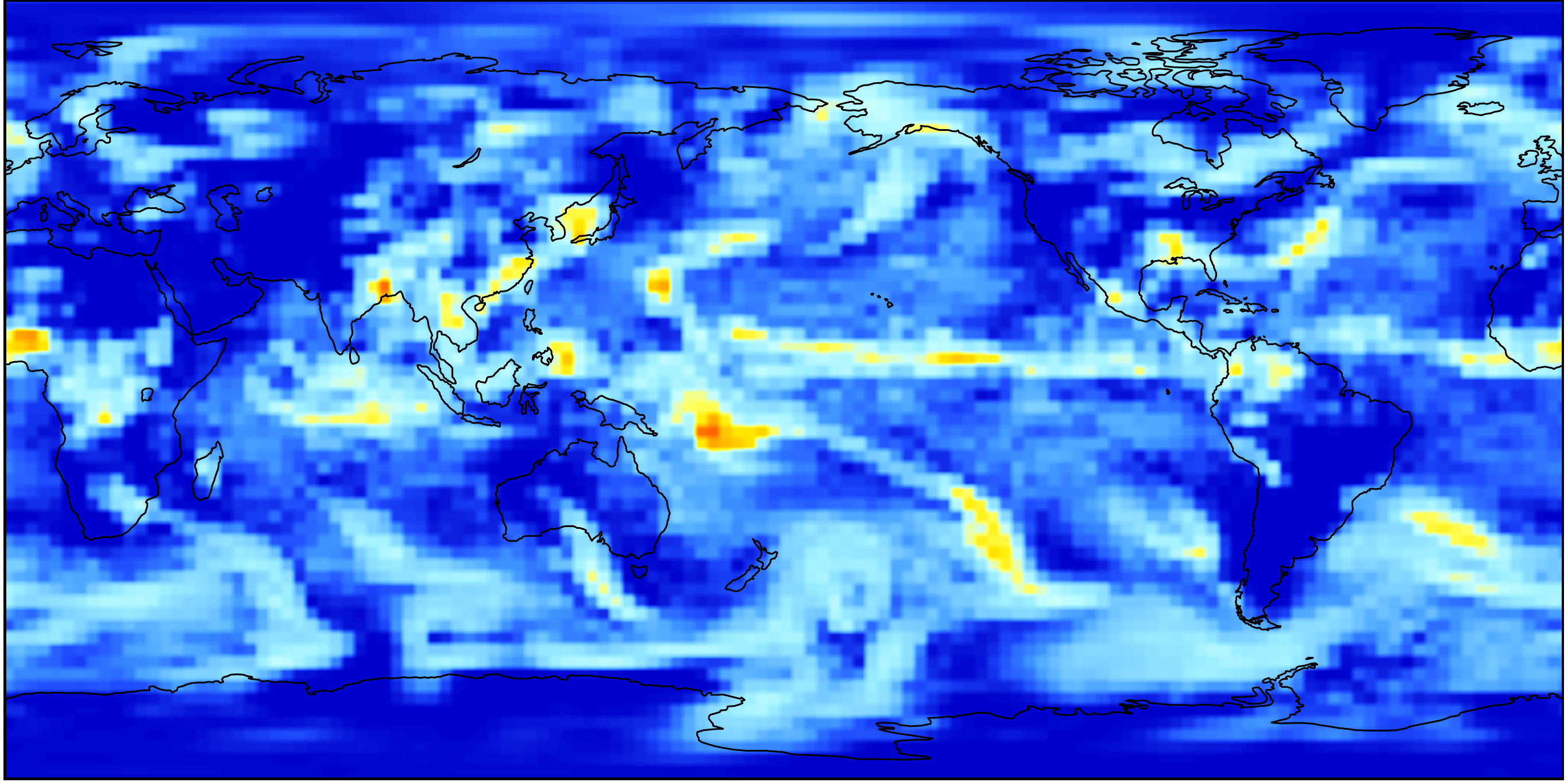}
    \includegraphics[width=0.45\textwidth]{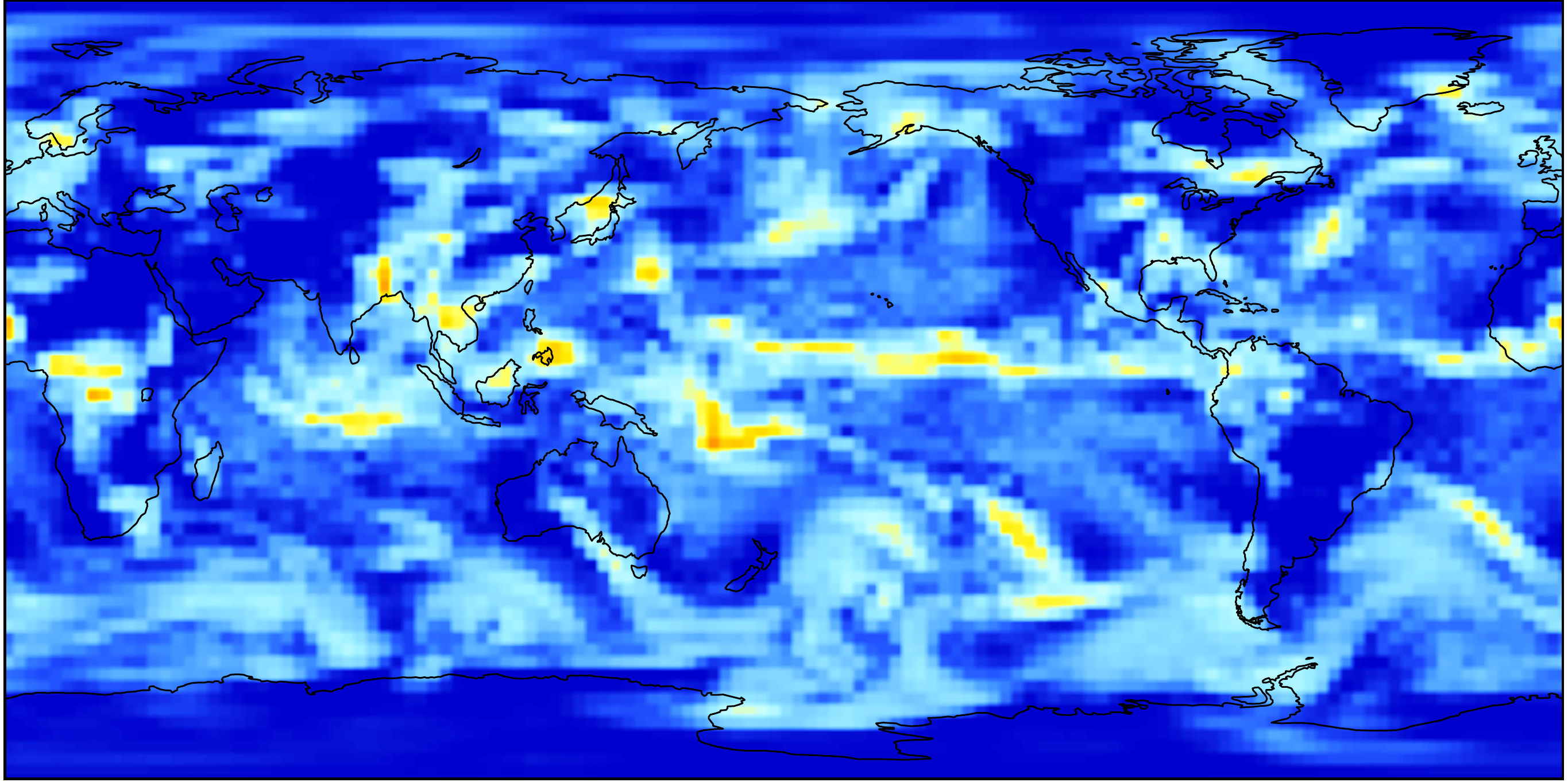}
    \caption{Precipitation maps for four successive test set days in fall and winter}
    \label{fig:real_fall_0}
\end{figure}

Figures \ref{fig:gen_spring_0}, \ref{fig:gen_fall_0}, \ref{fig:real_spring_0}, and \ref{fig:real_fall_0} are visualizations of log-normalized four-day sequences of the generated and test set precipitation data. Each sequence proceeds left-to-right, then top-to-bottom, and was randomly selected from a 32-day sample.

\end{document}